\documentclass[12pt,preprint]{aastex}
\usepackage{amsmath,amssymb,multirow}

\def\n{{\rm n}}
\def\p{{\rm p}}
\def\x{{\rm x}}

\def\bdxi{{\boldsymbol \xi}}
\def\bdeta{{\boldsymbol \eta}}
\def\bdr{{\boldsymbol r}}

\def\bdD{{\boldsymbol D}}
\def\bdU{{\boldsymbol U}}

\begin{document}

\title{Universality in oscillation modes of superfluid neutron stars?}

\author{K.~S. Wong, L.~M. Lin, and P.~T. Leung}
\affil{Department of Physics and Institute of Theoretical Physics, 
The Chinese University of Hong Kong, Shatin, Hong Kong SAR, China}

\date{\today}

\begin{abstract}  
It has been well established that the $f$-mode of relativistic
ordinary-fluid neutron stars displays a universal scaling behavior. 
Here we study whether the ``ordinary'' $f_{\rm o}$- and ``superfluid'' 
$f_{\rm s}$-modes of superfluid neutron stars also show similar universal 
behavior. 
We first consider a simple case where the neutron superfluid and normal
fluid are decoupled, and with each fluid modeled by a polytropic equation 
of state.
We find that the $f_{\rm o}$-mode obeys the same scaling 
laws as established for the $f$-mode of orindary-fluid stars. However, 
the oscillation frequency of the $f_{\rm s}$-mode obeys a 
different scaling law, which can be derived analytically from a 
homogenous two-fluid stellar model in Newtonian gravity. Next the coupling 
effect between the two fluids is studied via a parameterized
model of entrainment. We find that the coupling in general breaks the universal
behavior seen in the case of decoupled fluids. Based on a relativistic 
variational principle, an approximated expression is derived for the 
first-order shift of the $f_{\rm s}$-mode squared frequency due to the 
entrainment.

\end{abstract}


\section{Introduction}
\label{sec:intro}

Ground-based gravitational wave detectors are either already operating or 
will soon be operating. For example, the LIGO detectors have conducted five
science runs since 2002. While the current detectors are still not sensitive 
enough to detect gravitational waves directly, interesting upper limits on 
the gravitational-wave strains from several potential astrophysical sources 
(e.g., isolated pulsars and stochastic background) have been placed
\citep{abbott-2007-1,abbott-2007-2}.
It is very likely that the detection of gravitational waves will come within a 
decade or so after the upgrade of the detectors.

Pulsating neutron stars are one of the interesting potential sources of 
gravitational waves. They may also rotate 
rapidly enough for various kinds of rotational induced mode instabilities to 
develop and enhance the emitted gravitational waves \citep{andersson2003gwi}.
The pulsation modes of neutron stars are damped due to the emission of 
gravitational waves, and hence are called quasinormal modes 
\citep[see, e.g.,][for reviews]{kokkotas1999qnm,ferrari08},
instead of normal modes as in Newtonian theory. The study of quasinormal 
modes of neutron stars has a long history dating back to the pioneering works 
of Thorne and his collaborators \citep{thorne1967tnr,price1969tnr,thorne1969npg1,thorne1969npg2,campolattaro1970npg}. 
It is by now well established that the quasinormal mode spectrum of a neutron 
star is tremendously rich and the gravitational waves emitted from a 
pulsating star carry important information about the internal structure of 
the star 
\citep[e.g.,][]{andersson96, andersson1998tgw,Benhar:1998au,kokkotas01,
benhar04,tsui2005prl}. 
The gravitational-wave signals from pulsating neutron stars, should they be 
detected by ongoing or future detectors, can thus in principle be used to 
constrain the supranuclear equation of state (EOS), which is still 
poorly understood.

While the quasinormal modes of a pulsating neutron star in general depend 
sensitively on the stellar model and EOS, some empirical universal behaviors 
have also been observed with different EOSs. \citet{andersson1998tgw}  
noted that the frequencies and damping times 
of the leading gravitational-wave $w$-modes and fluid $f$-mode of nonrotating
neutron stars can be approximated by some empirical relations which depend 
only on the mass and radius of the star for most EOSs
\citep[see also][]{Benhar:1998au}.
In particular, \citet{andersson1998tgw} obtained the following formulas 
respectively for the real and imaginary parts of $f$-mode frequency: 
\begin{equation}
{\rm Re}(\omega M) = \alpha_1 M + \alpha_2 C^{3/2} , 
\label{eq:nils_real_f}
\end{equation}
\begin{equation}
{\rm Im}(\omega M) =  C^4 \left( \beta_1 C + \beta_2 \right) , 
\label{eq:nils_img_f} 
\end{equation} 
where $\alpha_1$, $\alpha_2$, $\beta_1$, and $\beta_2$ are model-independent
constants determined by curve fitting. $M$ is the mass of the star and
the dimensionless parameter $C \equiv M/R$ (in units where $G=c=1$) is the 
compactness of the star. 
The real part of $\omega$ is the mode oscillation frequency, while the 
imaginary part is inversely proportional to the damping time due to the 
emission of gravitational waves.   
Recently, the physical mechanism behind such universal behaviors has been 
investigated in detail by \citet{tsui2005uqn}. 
In their notations, the (complex) frequency $\omega $ of 
the $w$-mode or $f$-mode can be approximated by 
\begin{equation}
\omega M = a C^2 + b C + c ,
\label{eq:scaling_law}
\end{equation}
where the complex parameters $(a,b,c)$ are also model-independent constants 
determined by curve fitting. 
Exploring such universality for the $w$-modes, 
\citet{tsui2005prl} have developed an inversion scheme to determine the 
mass, radius, density distribution, and the EOS of a neutron 
star model from the frequencies of the first few $w$-modes of the star.

Mature neutron stars are known to be very cold on the nuclear temperature
scale ($10^{10}$ K). It is believed that a newborn neutron star can 
cool down rather quickly (on a timescale of a few weeks to months) to the 
transition temperature ($\sim 10^9$ K) for nucleon superfluidity and 
superconductivity to occur 
\citep[see, e.g.,][]{lombardo-1999,lombardo-2001-578,andersson-2005-763}.
For sufficiently high density, 
nuclear matter can transform to deconfined quark matter, which may also 
be in the so-called color-superconducting phase 
\citep[see, e.g.,][]{alford-2003-67,alford-2004-30}.
To date, there 
is still no direct evidence for the existence of nucleon superfluidity in 
neutron stars. However, the well-established pulsar glitch phenomenon 
is best explained by the pinning and unpinning of large numbers of superfluid
vortices to the solid crust \citep{Radhakrishnan-1969,lyne-1993}. 
So, how would nucleon superfluidity 
affect the dynamics and hence the oscillation mode spectrum of neutron stars?  
For the simplest model, neutron stars consist of neutrons, protons, and 
electrons. When neutrons become superfluid, they dynamically decouple 
from protons and electrons. 
The protons can also be in a superconducting state, but
they are coupled to the ordinary electron fluid via electromagnetic intereaction on a very short timescale. In effect, the neutron star interior can 
be approximated by a two-fluid model: one fluid is the neutron superfluid; the
other fluid is a conglomerate of all other charged constituents.

To a first approximation, the two interpenetrating fluids could be considered
as independent. However, due to the strong interaction between neutrons and 
protons, the two fluids in general can couple via the so-called entrainment 
effect. Entrainment is a multifluid effect and it arises when the flow of 
one fluid induces a momentum in the other fluid. 
In this work, we will study the parameterized entrainment model 
used previously by \citet{andersson-2002-66}.

Building on the formalism developed by Carter and his collaborators
\citep[e.g.,][]{carter1989rfd,comer1993hfm,comer1994hfr,carter-1995-454,
langlois1998drr}, \citet{comer-1999-60} and \citet{andersson-2002-66} 
have calculated the quasinormal modes of such nonrotating two-fluid stellar 
model 
\citep[see, e.g.,][for the corresponding study in Newtonian theory]{lindblom1994osn,lee1995non,prix2002aon}. 
These works show the existence of a new family of modes for a two-fluid star, 
the so-called superfluid modes. These modes have the distinguishing 
characteristic that the two fluids are essentially counter-moving, as a 
manifestation of an extra fluid degree of freedom which is missing in 
the study of pulsating ordinary, single fluid neutron stars 
\citep{comer-1999-60}.  
Another characteristic of the superfluid modes is that they depend 
sensitively on the entrainment effects between the two fluids
\citep{andersson-2002-66}. 
In a two-fluid star, there is essentially a doubling of the fluid modes. 
For example, the single $f$-mode which exists in an ordinary fluid 
star is splitted into an ``ordinary'' $f$-mode (denoted by $f_{\rm o}$),
where the two fluids tend to move together, and a superfluid counterpart 
(denoted by $f_{\rm s}$) where the fluids are counter-moving 
\citep{comer-1999-60}.  
On the other hand, the $w$-modes do not show such kind of mode 
doubling effect \citep{comer-1999-60}. The two fluids of the $w$-modes 
always move in ``lock-step'' due to the fact that $w$-modes are largely 
spacetime oscillations. 

As mentioned above, the frequencies of the $f$- (or $w$-modes) of 
nonrotating ordinary-fluid neutron stars can be approximated by 
universal scaling laws (\ref{eq:nils_real_f})-(\ref{eq:scaling_law}) for 
different EOSs. 
It is thus natural to ask whether this universality also holds for 
two-fluid neutron star models. In this paper, we investigate whether the 
$f_{\rm o}$ and $f_{\rm s}$ modes of two-fluid stars establish any kind of 
universality. 
In order to provide the reader a better understanding of the main 
results obtained in this paper, we shall give a brief summary in the following.

In this work we first study the simplest case of two decoupled fluids,
each with a polytropic EOS. We vary the polytropic indices of 
the two fluids to mimic the effects of different EOSs. 
We find that the $f_{\rm o}$-mode (i.e., the ``ordinary'' $f$-mode) can 
still be approximated very well by the same scaling laws 
(\ref{eq:nils_real_f})-(\ref{eq:scaling_law}) as established for 
ordinary-fluid neutron stars. 
For the superfluid $f_{\rm s}$-mode, we find that the real part of the mode
frequency satisfies a different universal scaling law 
(Eq.~(\ref{eq:newton_fs})), while the imaginary part in general does not 
have any universal behavior. We are also able to derive the universal scaling
curve for the real part of the $f_{\rm s}$-mode frequency analytically based 
on a homogeneous two-fluid model in Newtonian gravity. 

We then study the effect of coupling between the two fluids via a parameterized
model of entrainment. We analyze both numerically and analytically how the 
entrainment affects the $f_{\rm s}$-mode frequency. Based on a 
relativistic variational principle, we have derived a general integral 
formula to calculate the first-order shift in the mode (squared) frequency 
(Eq.~(\ref{eq:1st_order_rel})). 
Furthermore, for the particular class of background EOS and entrainment 
models we considered in the study, we are able to approximate
the integral formula by an algebraic expression (\ref{eq:1st_order_rel_final})
which depends explicitly only on the model parameters.

The main results obtained in this work
(Eqs.~(\ref{eq:newton_fs}) and (\ref{eq:w2_over_w2f2}))
can be used by other researchers to obtain a good approximation to the 
$f_{\rm s}$-mode frequency (without the need of constructing their own 
numerical code) for the class of background EOS and 
entrainment models that have been used extensively to study superfluid 
neutron stars.

The plan of this paper is as follows. Section~\ref{sec:formal} 
summarizes briefly the general relativistic two-fluid formalism. 
In section~\ref{sec:eos} we describe the EOS models we use in 
the study. Section~\ref{sec:results} presents our numerical results. 
Section~\ref{sec:fs_mode_derive} presents our analytical analysis for a
homogeneous two-fluid star model. 
The effects of entrainment on the superfluid mode frequency is studied 
analytically in Section~\ref{sec:pert}.  
Finally, we summarize our results in section~\ref{sec:conclude}. 
We use units where $G=c=1$ unless otherwise noted.

\section{The Two-Fluid Formalism}
\label{sec:formal}

In this paper we make use of the formalism and numerical code developed by 
\citet{comer-1999-60} to compute the oscillation modes of a general 
relativistic superfluid star. 
As discussed in Sec.~\ref{sec:intro}, the superfluid neutron star interior is 
approximated by a two-fluid model. 
One of the fluid is composed of superfluid neutrons, and
the other fluid contains all other charged particles (proton, electrons,
and crust nuclei etc), which will be called proton fluid for simplicity. 
We focus our attention to a simplified model where the two fluids are assumed 
to coexist throughout the whole star as in \citet{comer-1999-60}. 
The work by \citet{comer-1999-60} is based on Carter's 
general relativistic superfluid formalism. Here we will only summarize the 
formalism briefly \citep[see][for a recent review]{Andersson_Comer_review}.

The central quantity of the two-fluid formalism is the master function 
$\Lambda(n^2,p^2,x^2)$, which is formed by three scalars, 
$n^2=-n_{\alpha}n^{\alpha}$, $p^2=-p_{\alpha}p^{\alpha}$, 
and $x^2=-n_{\alpha}p^{\alpha}$. 
The four vectors $n^{\alpha}$ and $p^{\alpha}$ are respectively the conserved 
number density currents for the neutrons and protons. The master function is 
a two-fluid analog of the EOS. Here we take $-\Lambda$ to be 
the total thermodynamic energy density. Given a master function
$\Lambda$, the stress-energy tensor is found to be 
\begin{equation}
T^{\alpha}_{\beta} = \Psi \delta^{\alpha}_{\beta} + n^{\alpha}\mu_{\beta}
 + p^{\alpha} \chi_{\beta} , 
\end{equation}
where $\Psi$ is the generalized pressure 
\begin{equation}
\Psi = \Lambda - n^{\alpha}\mu_{\alpha} - p^{\alpha}\chi_{\alpha} . 
\end{equation}
The chemical potential covectors $\mu_{\alpha}$ and $\chi_{\alpha}$, 
respectively, for neutrons and (conglomerate) protons are given by 
\begin{equation}
\mu_{\alpha} = {\cal B} n_{\alpha} + {\cal A} p_{\alpha} , \ \ 
\chi_{\alpha} = {\cal C} p_{\alpha} + {\cal A} n_{\alpha} ,
\end{equation}
where 
\begin{equation}
{\cal A} = -{\partial \Lambda\over \partial x^2}, \ \ 
{\cal B} = -2 {\partial \Lambda\over \partial n^2}, \ \ 
{\cal C} = -2 {\partial \Lambda\over \partial p^2} . 
\label{eq:abc_coeff}
\end{equation}
It is noted that the so-called entrainment effect is described by 
the coefficient ${\cal A}$. If ${\cal A} \neq 0$, the mass 
current of one particle constituent will induce a momentum in the other
constituent.

The equations of motion for the two-fluid system consist of two 
conservation equations, 
\begin{equation}
\nabla_{\alpha} n^{\alpha} = 0 , \ \ \nabla_{\alpha} p^{\alpha} = 0 , 
\label{eq:np_con}
\end{equation}
and two Euler equations, 
\begin{equation}
n^{\alpha} \nabla_{[\alpha }\mu_{\beta ] } = 0 , \ \ 
p^{\alpha} \nabla_{[\alpha } \chi_{\beta ]} = 0 .
\label{eq:euler}
\end{equation}
Solving the linearized superfluid and Einstein field equations, 
\citet{comer-1999-60} obtained a system of first order 
differential equations to describe the (even-parity or polar) non-radial 
oscillations of superfluid neutron stars. 
We refer the readers to \citet{comer-1999-60} for the 
detailed equations and numerical techniques.

\section{Equation of State}
\label{sec:eos}

As discussed in the previous section, the master function 
$\Lambda(n^2,p^2,x^2)$ is a two-fluid analog of the EOS, replacing the 
role of $P=P(\rho)$ (with $P$ and $\rho$ being respectively 
the pressure and total energy density) for ordinary one-fluid stellar 
models. While the universality of the $f$-mode of ordinary fluid neutron
stars is determined by many different EOSs, we cannot employ those EOSs 
directly for our two-fluid neutron stars. 
The reason is that realistic nuclear-matter EOSs are generally given in the 
form of pressure vs density. This is clearly insufficient for the two-fluid 
formalism used here in which the energy density $-\Lambda$ must be given 
as a function of the two number densities $(n,p)$, and hence standard 
tabulated EOSs are incomplete for our purposes. 
We refer the reader to \citet{lin-2007} for a detailed discussion on 
what is required from next generation EOSs that allow for proper treatment of 
the dynamics of multi-fluid neutron stars.

In this paper, we shall first study the oscillation modes of superfluid 
neutron stars without considering the entrainment. The master 
function $\Lambda$ is taken to be
\begin{equation}
\Lambda = \lambda_0 \equiv 
- m (n + p) - \sigma_{\n} n^{\beta_{\n}} - \sigma_{\p} p^{\beta_{\p}} ,
\label{eq:master_poly}
\end{equation}
where the two fluid particles are assumed to have the same mass $m$
(i.e., the baryon mass). The parameters $\sigma_{\n}$, $\sigma_{\p}$, 
$\beta_{\n}$ and $\beta_{\p}$ are freely chosen. This master function has been
used extensively to study superfluid neutron stars
\citep{comer-1999-60,andersson2001srg,andersson-2002-66,prix2002aon,
yoshida03}.
For this EOS, each fluid behaves as a relativistic polytrope and can be 
regarded as decoupled (though it should be noted that the 
fluids still couple globally through gravity). 
In general, we say that the two fluids are decoupled if the master function 
is separable, in the sense that it can be decomposed into two contributions 
corresponding respectively to each fluid, i.e., 
$\Lambda(n^2,p^2)=\Lambda_{\n}(n^2)+\Lambda_{\p}(p^2)$.

\subsection{Entrainment}
\label{Entrainment}

After studying the decoupled case, we shall also study the entrainment 
between the fluids. 
As already mentioned, entrainment describes essentially the effect where the momentum of one species depends on the relative motion between the two species. In the relativistic framework, this effect is specified by the coefficient $ \mathcal{A} $, which is in turn determined by the dependence of master function $ \Lambda $ on $ x^2 $. Note that since the background is assumed static, $ x^2 -np $ should remain small even when the star oscillates. We shall follow 
\citet{andersson-2002-66} and approximate the master function by  
\begin{equation}
\Lambda = \lambda_0  + \lambda_1 (x^2 - n p) ,
\label{eq:master_lambda01}
\end{equation}
where the first term $\lambda_0$ is given by Eq. (\ref{eq:master_poly}). 
The second term is used to describe the entrainment and $\lambda_1$ is given 
by  
\begin{equation}
\lambda_1(n^2, p^2) = \epsilon \frac{m}{ p + \epsilon ( n + p)} .
\label{eq:master_lambda1}
\end{equation}
The positive constant $\epsilon$ is used to parametrize the strength of 
the entrainment. As discussed in \citet{andersson-2002-66},
we shall consider the physically reasonable range to be 
$0 \leq \epsilon \leq 0.2$. 
Note that the second term of the master function $\Lambda$ vanishes in the 
background (where $x^2 = n p$). Hence, the background quantities of the star
are determined solely by $\lambda_0$. 
We give the expressions for some of the thermodynamic coefficients 
on the background which will be used in later sections:
\begin{eqnarray}
\mathcal{A} &=& - \lambda_1 (n^2, p^2) ,
\nonumber\\
\mathcal{B} &=& - {1 \over n}{\partial \lambda_0 \over \partial n} - {p \over n} \mathcal{A} ,
\nonumber\\
\mathcal{C} &=& - {1 \over p}{\partial \lambda_0 \over \partial p} - {n \over p} \mathcal{A} .
\label{eq:thm_coeff}
\end{eqnarray}

\section{Numerical Results}
\label{sec:results}

\subsection{Decoupled case} 
\label{sec:num_decouple}

In order to study whether the $f_{\rm o}$- and $f_{\rm s}$-modes of superfluid
neutron stars establish any kind of universality, we have varied the values 
of $\beta_{\n}$ and $\beta_{\p}$ for the master function 
(Eq. (\ref{eq:master_poly}))
to mimic the effects of different EOSs. 
The effects of $\sigma_{\n}$ and $\sigma_{\p}$ on the mode frequencies are 
relatively small and hence we shall fix them to be 
$\sigma_{\n}=\sigma_{\p}=0.5 m$. Table~\ref{tab:1}
summarizes the different EOS models we shall study in this section. 
In Table~\ref{tab:1} we have defined two parameters 
$\bar{\beta} = (\beta_{\n}+\beta_{\p})/2$ and $\Delta \beta=\beta_{\n}-\beta_{\p}$.
The parameter $\bar{\beta}$ can be considered as a parameter that controls 
the bulk fluid motion, while $\Delta \beta$ represents the asymmetry between 
the two fluids. 
A larger value of $\Delta \beta$ would imply that 
the effects of the two-fluid dynamics become more important. 
On the other hand, the case $\Delta \beta=0$ represents the one-fluid limit 
in which case the two fluid components are indistinguishable and the 
stellar models reduce to ordinary one-fluid neutron stars.

\begin{figure*}
\centering
  \begin{minipage}{0.4\linewidth}
        \centering
        \includegraphics*[width=6.5cm]{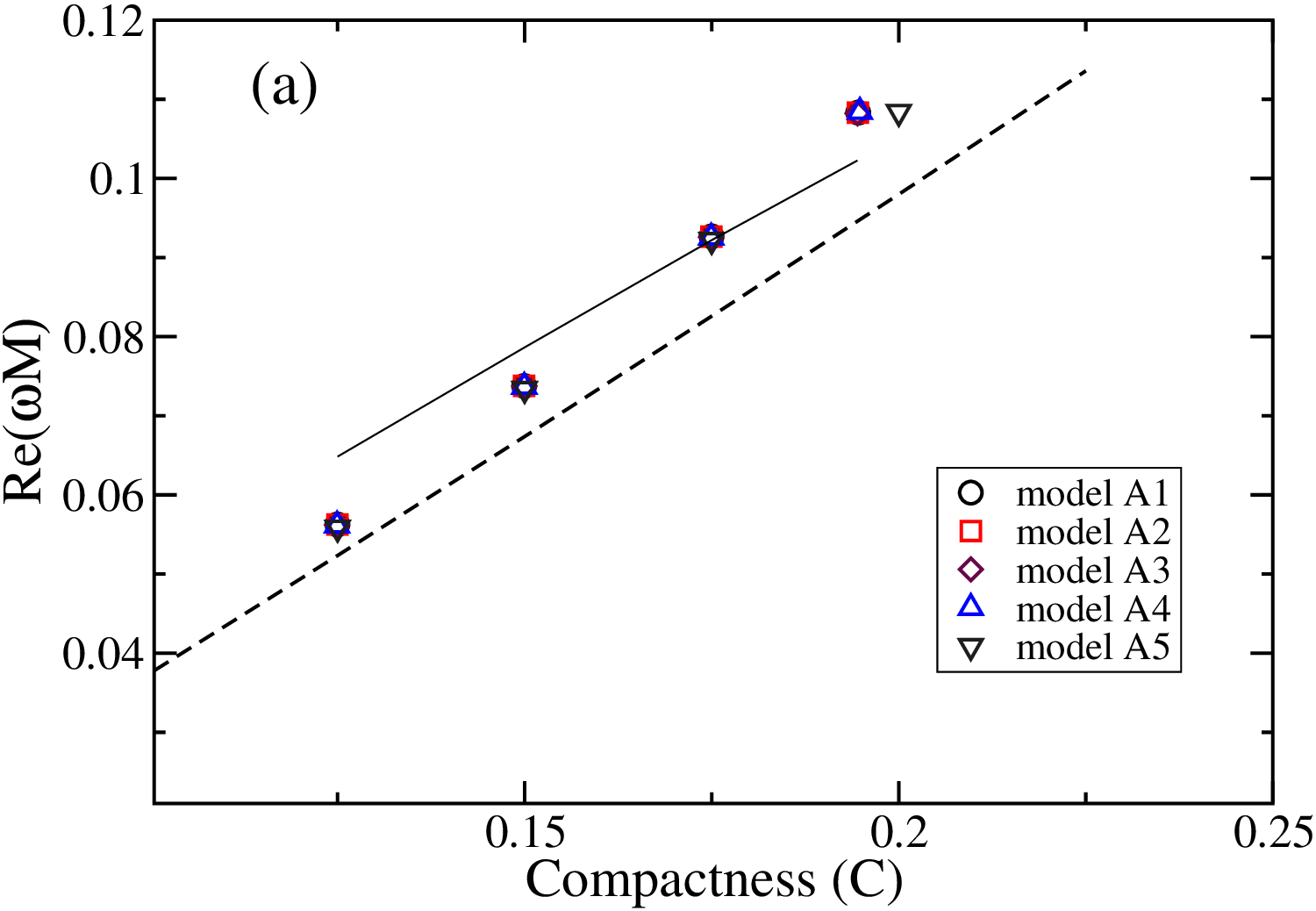}
  \end{minipage}%
  \begin{minipage}{0.4\linewidth}
        \centering
        \includegraphics*[width=6.5cm]{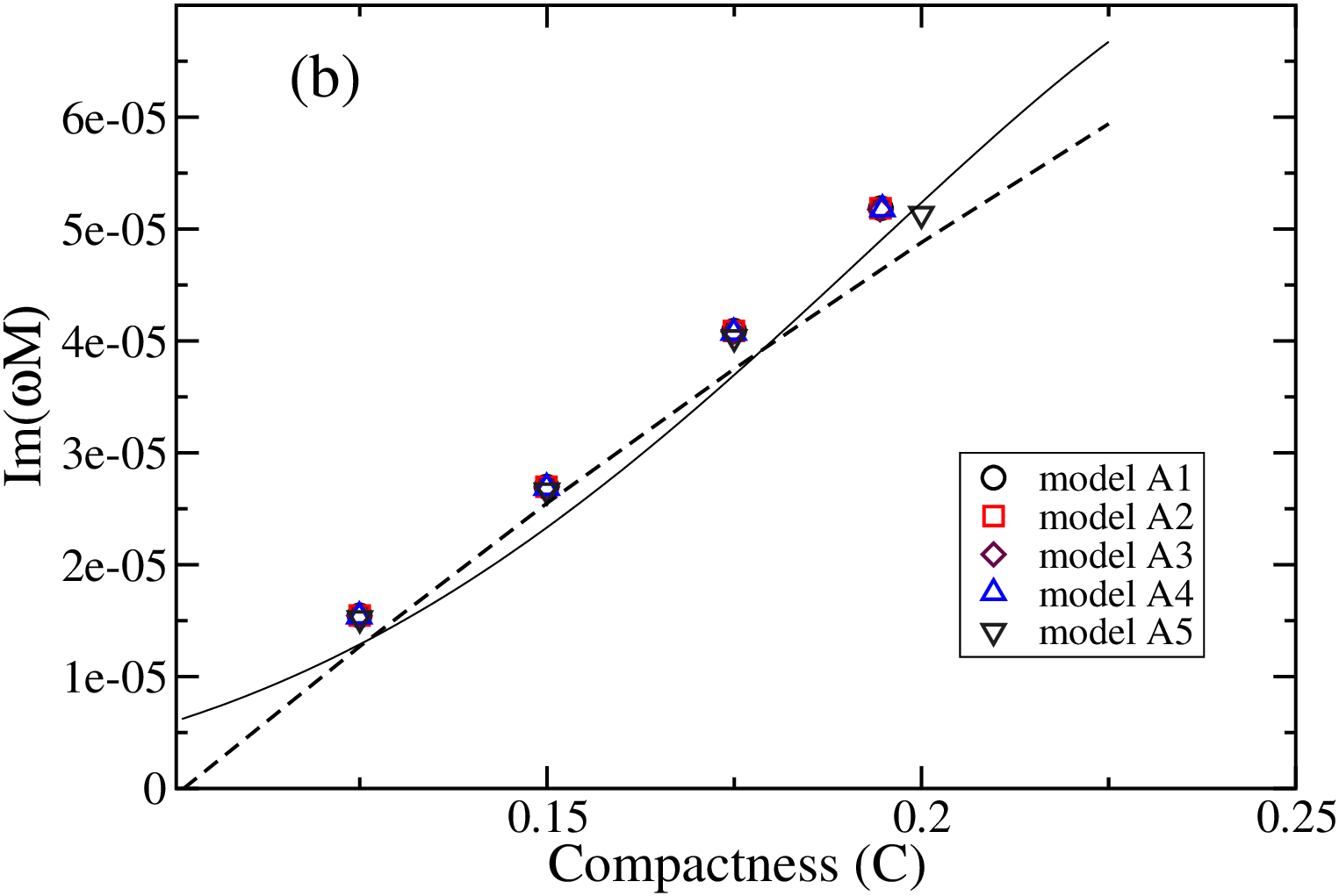}
  \end{minipage}
        \caption{(a) The real and (b) the imaginary parts of $\omega M$ of 
the $f_{\rm o}$-mode are plotted against the compactness $C$ for neutron stars 
described by the EOS models A$i\ (i=1-5)$. The solid line in (a) represents 
Eq.~(\ref{eq:nils_real_f}) for stars models with $M$ and $C$ obtained by 
EOS A1, while the solid line in (b) represents Eq.~(\ref{eq:nils_img_f}). 
The dashed lines in both figures represent Eq.~(\ref{eq:scaling_law}).}
        \label{fig:fo_A}
\end{figure*}

In Figures~\ref{fig:fo_A} (a) and (b) we plot the real and imaginary parts 
of the $f_{\rm o}$-mode frequencies respectively against the compactness 
$C=M/R$ for the EOS models A$i$ (with $i=1-5$). 
These models have the same ``bulk'' EOS parameter $\bar \beta=1.9$, but 
different ``asymmetry'' parameters $\Delta \beta$. 
In particular, Model A1 (with $\Delta \beta =0$) corresponds to the 
one-fluid limit for this series, while model A5 (with $\Delta \beta = 0.2$)
represents the most ``asymmetric'' one. 
For comparsion, we also plot the 
universal curves (Eqs.~(\ref{eq:nils_real_f})-(\ref{eq:scaling_law})) for the 
$f$-mode of ordinary-fluid neutron stars in the figures. 
In Figure~\ref{fig:fo_A} (a), the solid line represents 
Eq.~(\ref{eq:nils_real_f}) for the star models 
with the masses and compactnesses obtained by EOS A1, while the dashed line 
represents Eq.~(\ref{eq:scaling_law}). 
Similarly, in Figures~\ref{fig:fo_A} (b), the solid and 
dashed lines represent Eqs.~(\ref{eq:nils_img_f}) and (\ref{eq:scaling_law}) 
respectively. 
It is seen that the $f_{\rm o}$-modes of superfluid neutron stars still 
exhibit a universal behavior. For a given compactness $C$, 
the real and imaginary parts
of the mode frequency do not depend sensitively on the EOS. Furthermore, the 
mode frequency can still be approximated very well by the universal curves
(\ref{eq:nils_real_f})-(\ref{eq:scaling_law}) satisfied by the $f$-mode of 
ordinary-fluid neutron stars.

\begin{figure*}
\centering
  \begin{minipage}{0.4\linewidth}
        \centering
        \includegraphics*[width=6.5cm]{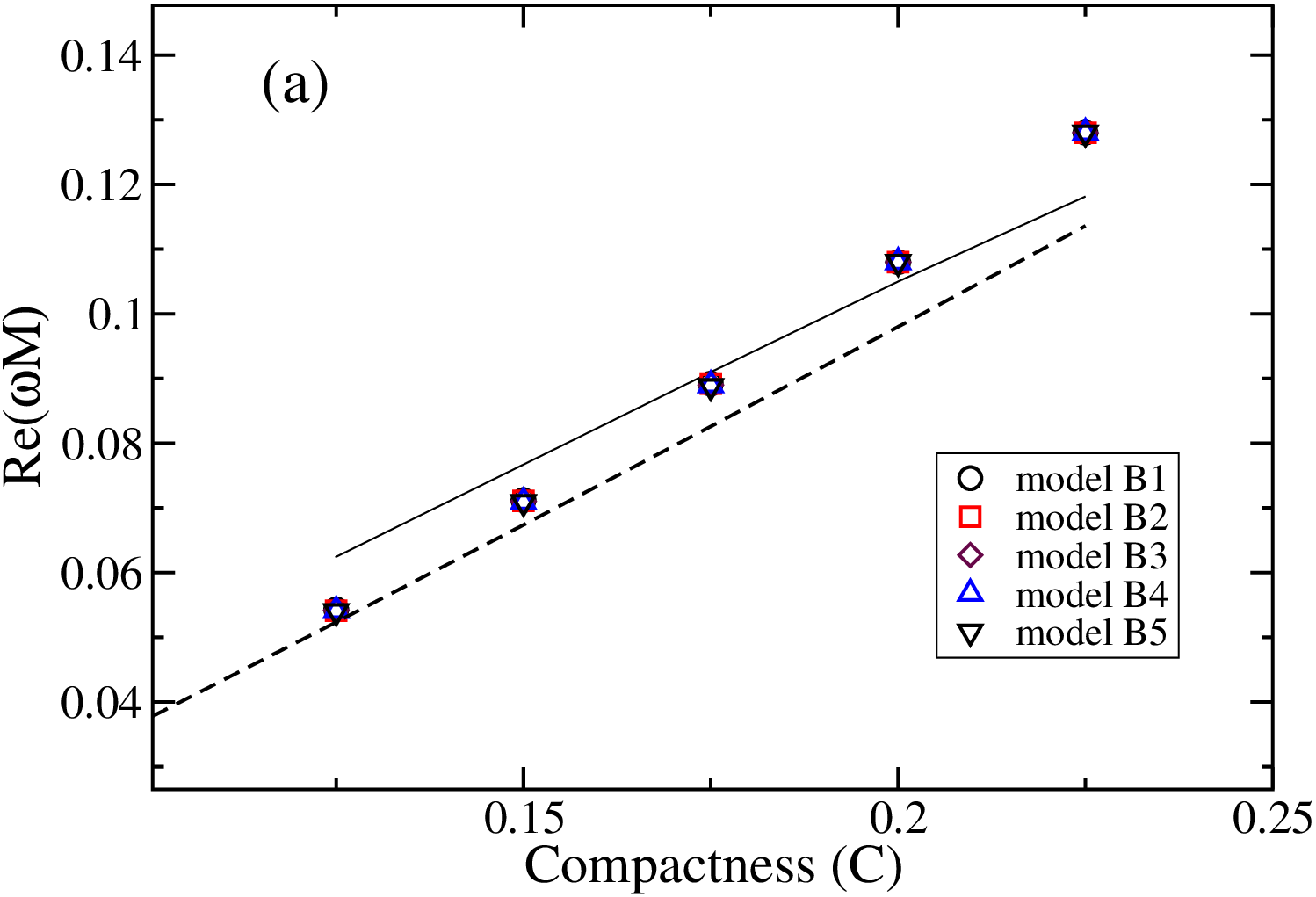}
  \end{minipage}%
  \begin{minipage}{0.4\linewidth}
        \centering
        \includegraphics*[width=6.5cm]{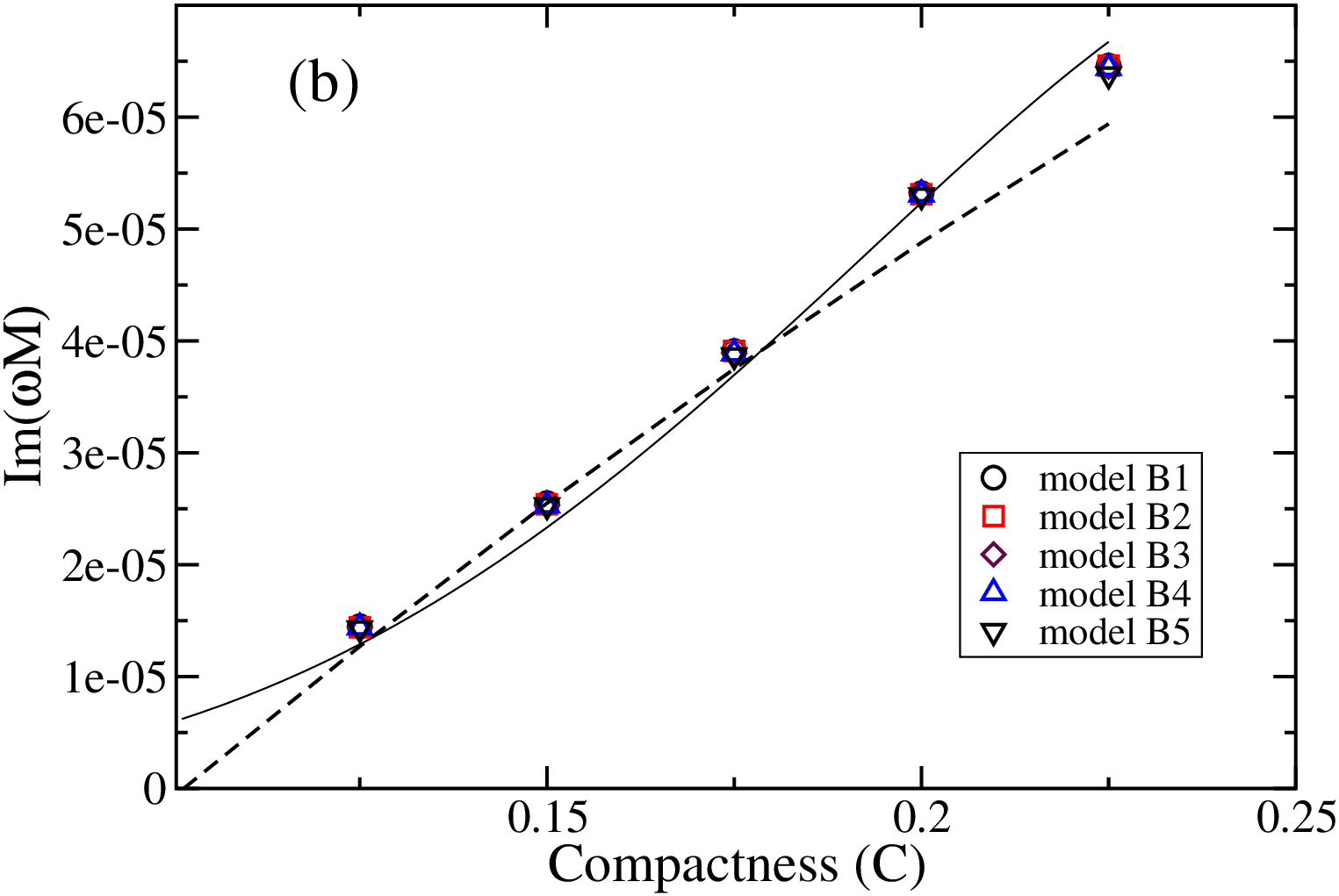}
  \end{minipage}
        \caption{Similar to Figure~\ref{fig:fo_A} but for the EOS models 
B$i\ (i=1-5)$. The solid line in (a) represents 
Eq.~(\ref{eq:nils_real_f}) for stars models with $M$ and $C$ obtained by 
EOS B1, while the solid line in (b) represents Eq.~(\ref{eq:nils_img_f}). 
The dashed lines in both figures represent Eq.~(\ref{eq:scaling_law}).
}
        \label{fig:fo_B}
\end{figure*}

\begin{figure*}
\centering
  \begin{minipage}{0.4\linewidth}
        \centering
        \includegraphics*[width=6.5cm]{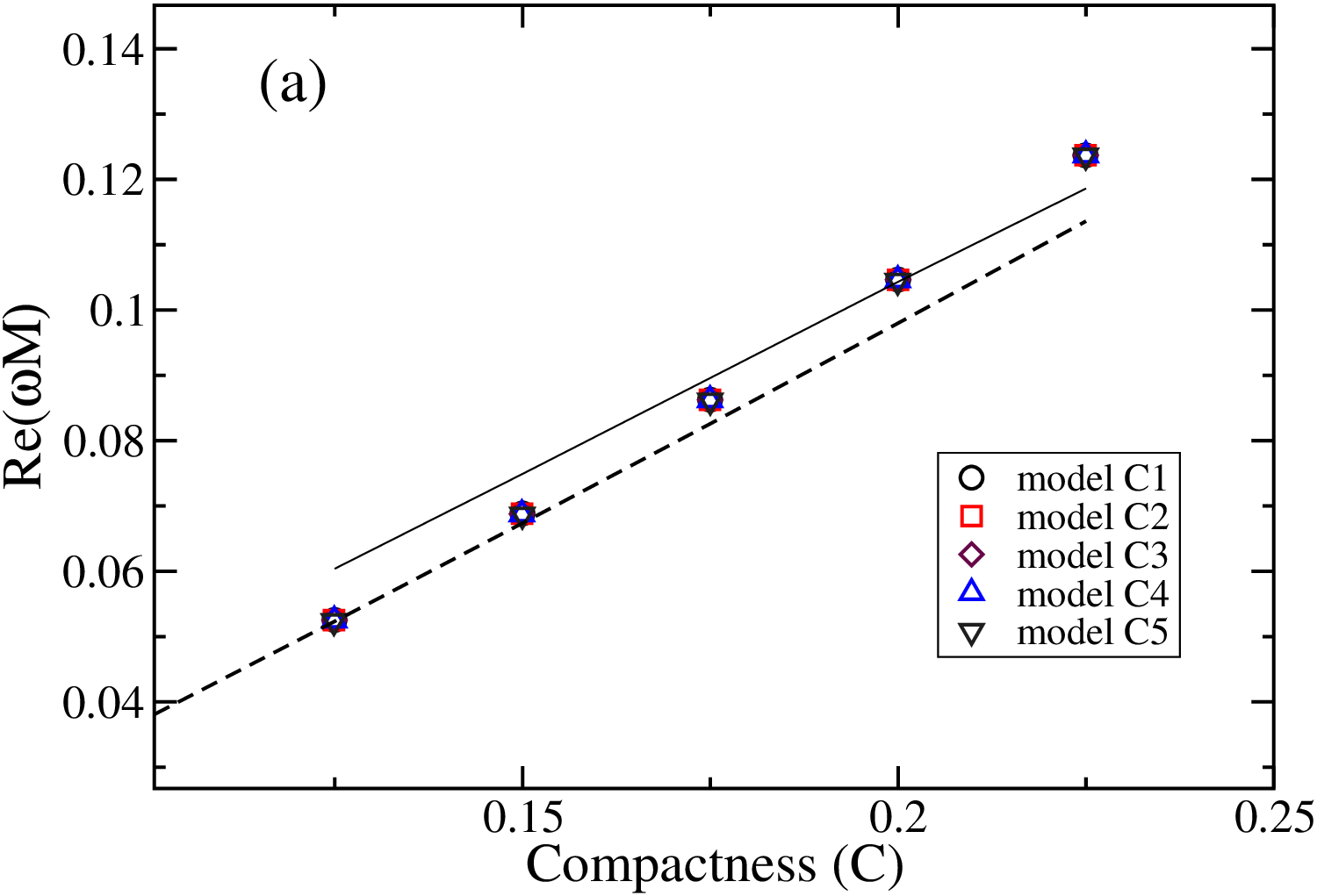}
  \end{minipage}%
  \begin{minipage}{0.4\linewidth}
        \centering
        \includegraphics*[width=6.5cm]{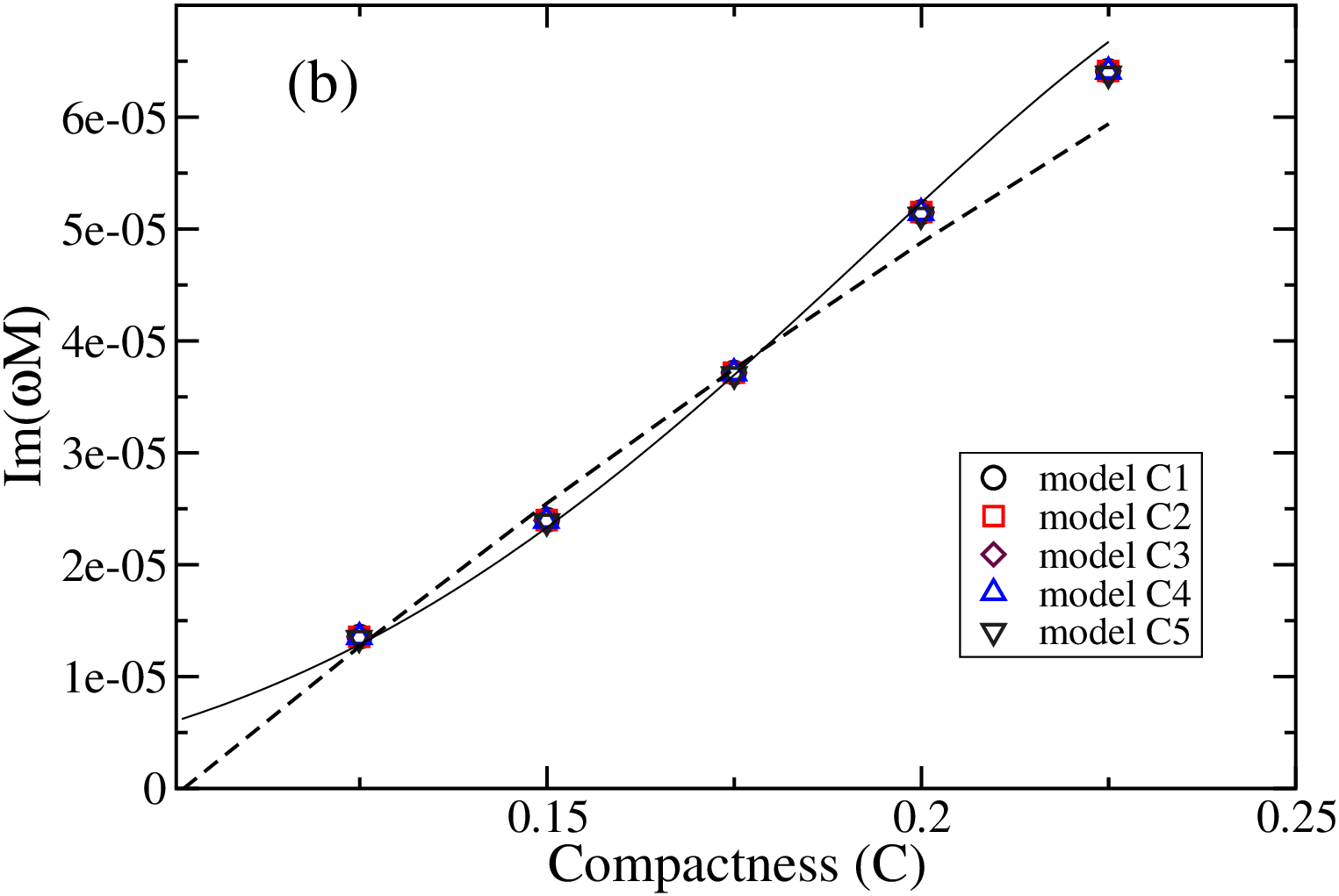}
  \end{minipage}
        \caption{Similar to Figure~\ref{fig:fo_A} but for the EOS models
C$i\ (i=1-5)$. The solid line in (a) represents 
Eq.~(\ref{eq:nils_real_f}) for stars models with $M$ and $C$ obtained by 
EOS C1, while the solid line in (b) represents Eq.~(\ref{eq:nils_img_f}). 
The dashed lines in both figures represent Eq.~(\ref{eq:scaling_law}).
}
        \label{fig:fo_C}
\end{figure*}

The corresponding results for the EOS models B$i$ and C$i$ (with $i=1-5$) are 
plotted in Figures~\ref{fig:fo_B} and \ref{fig:fo_C}. 
It is seen clearly from Figures~\ref{fig:fo_A} to \ref{fig:fo_C} that the 
$f_{\rm o}$-modes of superfluid neutron stars satisfy the same universal 
scaling as their ordinary-fluid counterpart. This result might not be so
suprising as the fluid motions of the $f_{\rm o}$-mode are such that 
the two fluids move in ``lock-step'' and reach essentially the same 
maximum displacements at the stellar surface: a feature that is similar to 
the situation of an ordinary-fluid neutron star in which the (single) fluid
displacement reaches a maximum at the surface 
\citep[see Fig.~9 of][]{comer-1999-60}.

\begin{figure}
\centering
\includegraphics*[width=6.8cm]{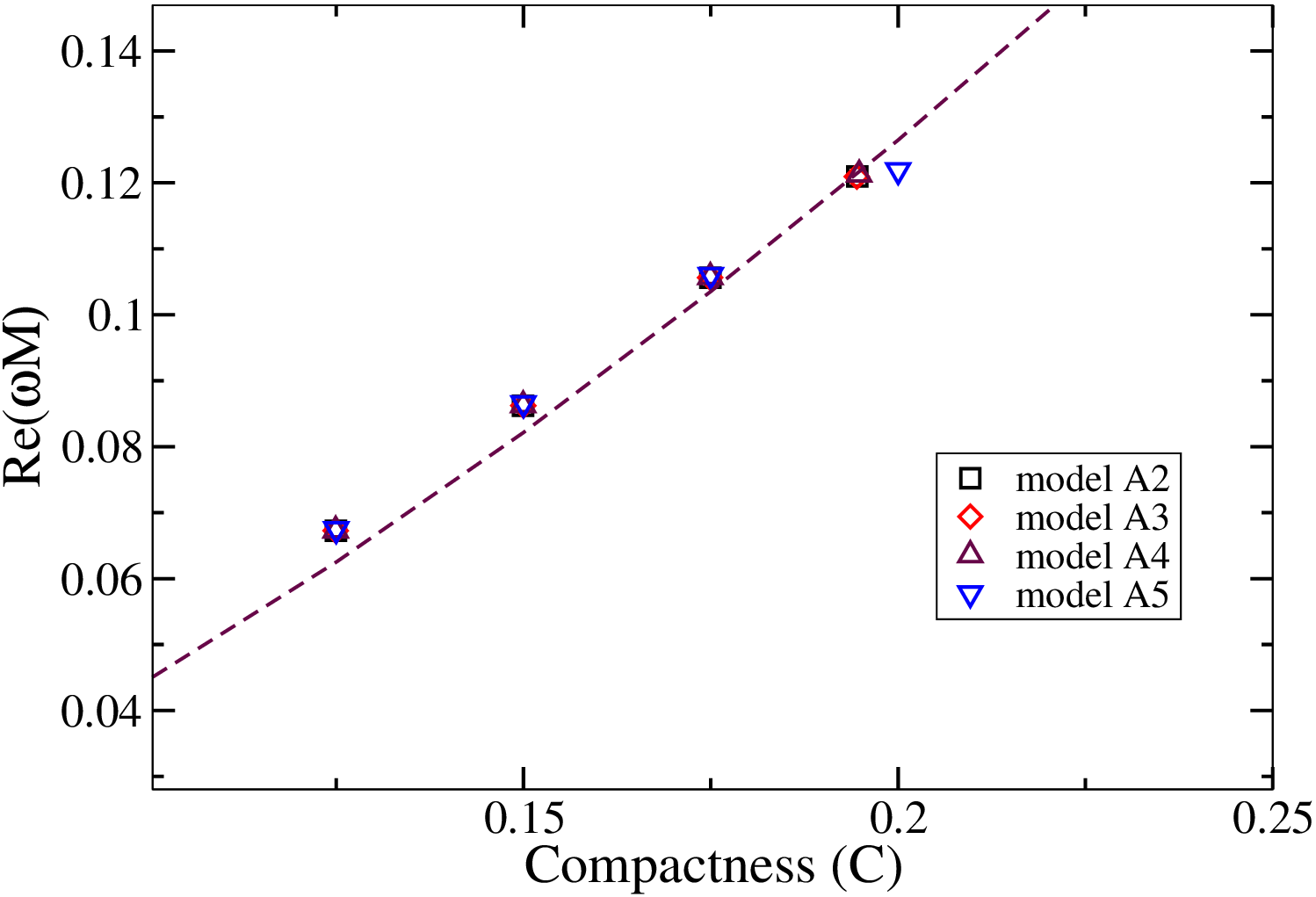}
\caption{
The real part of $\omega M$ of the $f_{\rm s}$-mode are plotted against 
the compactnesss $C$ for neutron stars described by the EOS models
A$i\ (i=2-5)$. The dashed line represents the analytic result 
(Eq.~(\ref{eq:newton_fs})). }
\label{fig:fs_RE_A}
\end{figure}

\begin{figure}
\centering
\includegraphics*[width=6.8cm]{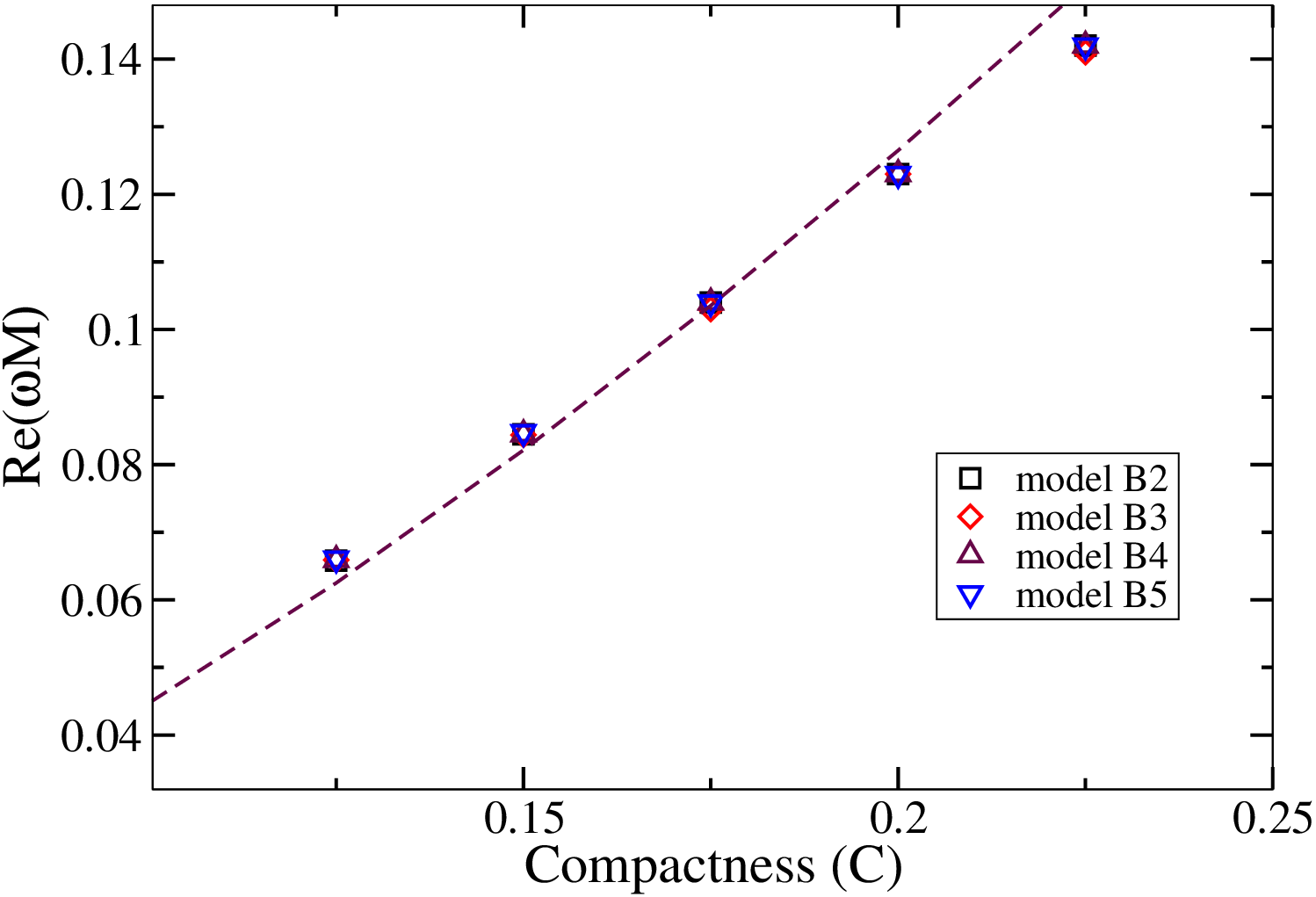}
\caption{Similar to Figure \ref{fig:fs_RE_A} but for the models 
B$i$ ($i=2-5$). }
\label{fig:fs_RE_B}
\end{figure}

\begin{figure}
\centering
\includegraphics*[width=6.8cm]{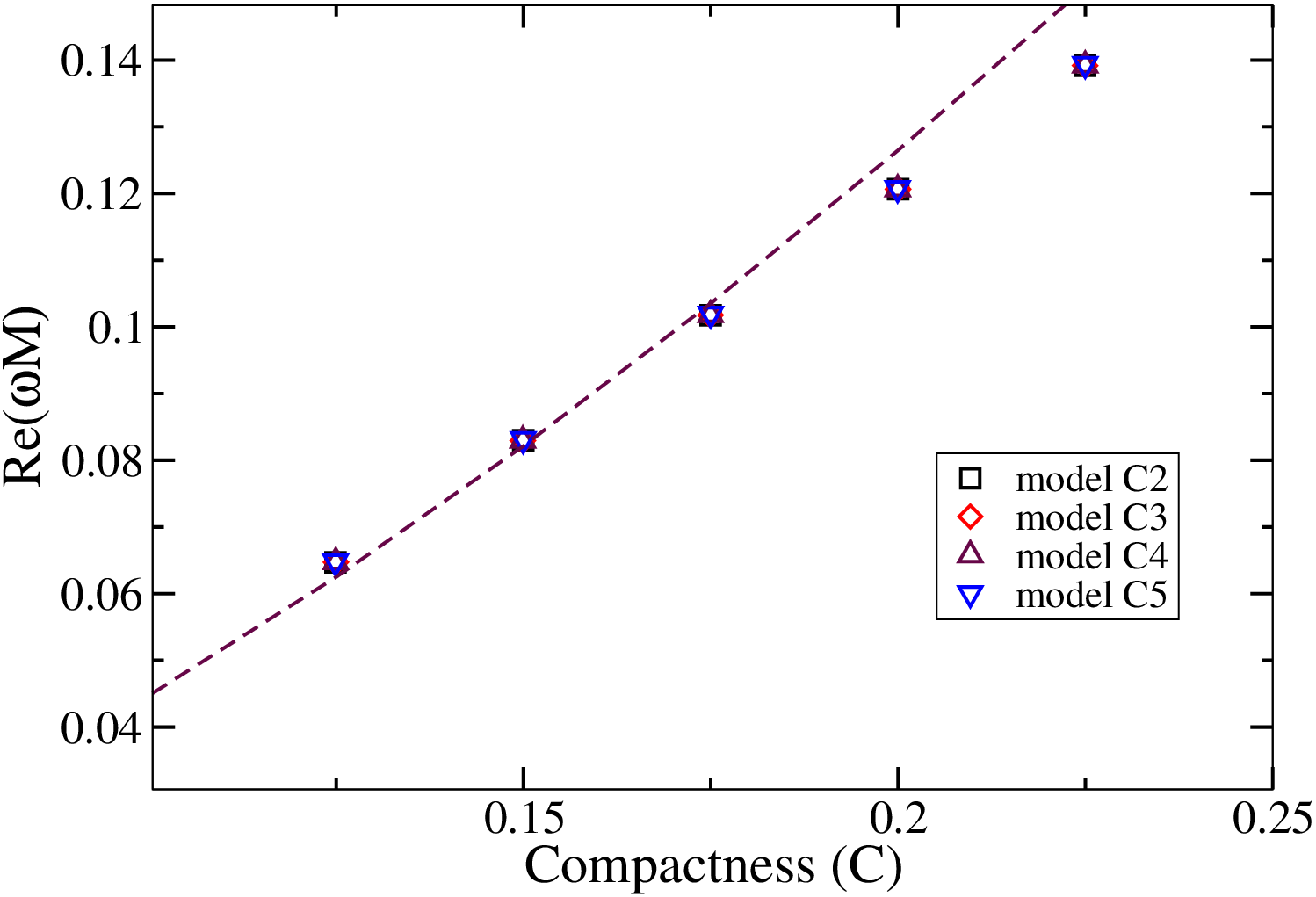}
\caption{Similar to Figure \ref{fig:fs_RE_A} but for the models C$i$
($i=2-5$).}
\label{fig:fs_RE_C}
\end{figure}

Now we consider the superfluid $f_{\rm s}$-mode, which does not have an 
ordinary-fluid counterpart. For this class of modes, there is no 
{\it a prior} reason to expect that they would establish a similar universal 
behavior. However, as shown in Figures~\ref{fig:fs_RE_A}, \ref{fig:fs_RE_B}, and \ref{fig:fs_RE_C}, the $f_{\rm s}$-modes do indeed follow 
another universal scaling law. In these figures, we plot the real parts of
the $f_{\rm s}$-mode frequencies against the compactness $C$ for the EOS models
A$i$, B$i$, and C$i$ (with $i=2-5$). 
In particular, the data can be approximated very well by the curve 
(dashed lines in the figures)
\begin{equation}
{\rm Re}(\omega M) = \left( l C^3 \right)^{1/2} , 
\label{eq:newton_fs}
\end{equation}
where $l$ is the spherical harmonics index. 
Note that the $f_{\rm s}$-mode does not exist in the models A1, B1, and C1 
since they correspond to the one-fluid limit $\Delta \beta=0$.

\begin{figure}
\centering
\includegraphics*[width=6.8cm]{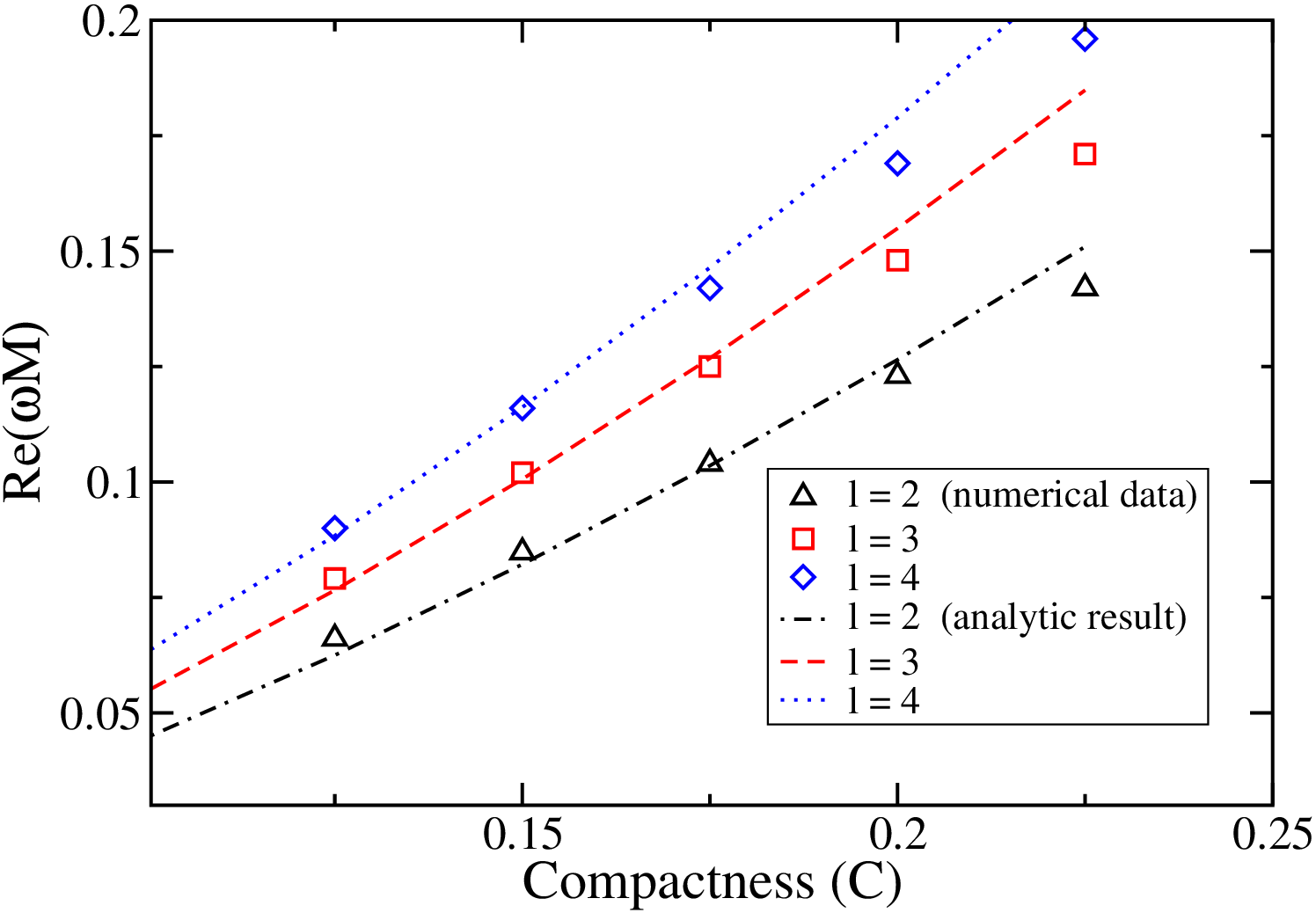}
\caption{Effects of the spherical harmonics index $l$: 
$\mathrm{Re}(\omega \mathrm{M})$ of the $f_{\rm s}$-modes 
vs $C$ for neutron stars described by the EOS model B5. The lines represent
Eq.~(\ref{eq:newton_fs}) for different values of $l$. }
\label{fig:fs_RE_l}
\end{figure}

In Figure~\ref{fig:fs_RE_l} we focus on the model B5 and plot the real part of 
the $f_{\rm s}$-mode frequency against the compactness for different values 
of the spherical harmonics index $l=2,3,4$. It is seen that the data can be 
approximated by Eq.~(\ref{eq:newton_fs}) very well in general. 
The discrepancy between the numerical data and Eq.~(\ref{eq:newton_fs}) 
becomes significant only for the $l=4$ modes in the region of high 
compactness ($C \gtrsim 0.2$). In Sec.~\ref{sec:fs_mode_derive}, we shall 
derive this scaling relation in the framework 
of Newtonian gravity and under the assumptions that the star is composed of 
two homogeneous and decoupled fluids. 
It should be noted that Eq.~(\ref{eq:newton_fs}) does not contain any 
free parameters, while the scaling equations 
(\ref{eq:nils_real_f})-(\ref{eq:scaling_law}) for the $f$-modes of 
ordinary-fluid stars depend on parameters that are determined by curve 
fitting \citep{andersson1998tgw,tsui2005uqn}.

It is also instructive to compare the scaling law (\ref{eq:nils_real_f}) for
the $f$-mode of ordinary-fluid stars with Eq.~(\ref{eq:newton_fs}) 
since they both contain the factor $C^{3/2}$. 
As discussed by \citet{andersson1998tgw}, since the characteristic timescale 
is related to the mean density of the star $\bar \rho$, it is expected that 
the $f$-mode frequency scales with $\bar{\rho}^{1/2} \sim (M/R^3)^{1/2}$. 
Fitting to the numerical data for the $f$-mode frequencies of ordinary-fluid
stars, \citet{andersson1998tgw} obtained the scaling law 
(\ref{eq:nils_real_f}). 
While Eqs.~(\ref{eq:nils_real_f}) and (\ref{eq:newton_fs}) both contain 
the scale factor $C^{3/2}$, we notice two important differences: (i) 
our scaling law (\ref{eq:newton_fs}) for the $f_{\rm s}$-modes does not 
contain any free parameter; (ii) it also does not contain a term that scales 
only with $M$ as in Eq.~(\ref{eq:nils_real_f}). 
As we shall show in Sec.~\ref{sec:fs_mode_derive}, the scaling law 
(\ref{eq:newton_fs}) is not a trivial generalization of 
Eq.~(\ref{eq:nils_real_f}) to the case of two-fluid stars. It is also not 
the well-known Kelvin mode for an incompressible stellar model 
\citep{chandra81}, whose oscillation frequency is given by 
\begin{equation}
\omega M = \left[ {{2 l (l-1) }\over {(2 l+1)} } C^3 \right]^{1/2} . 
\label{eq:kelvin_mode}
\end{equation}
However, Eqs.~(\ref{eq:newton_fs}) and (\ref{eq:kelvin_mode}) agree with each
other in the limit $l \gg 1$.

Now we turn to the imaginary part of the $f_{\rm s}$-mode. We see that in 
general it does not follow any universal scaling law. For a given compactness,
its value depend very sensitively on the EOS models. 
In Figure~\ref{fig:B_fs_IM} we plot ${\rm Im}(\omega M)$ against $C$ for 
models B$i$ ($i=2-5$). These models have the same ``bulk motion'' parameter
$\bar{\beta}=2$, but different ``asymmetry'' parameter $\Delta \beta$. 
It is seen that ${\rm Im}(\omega M)$ decreases significantly with 
decreasing $\Delta \beta$. In fact, for the particular form of 
the master function equation~(\ref{eq:master_poly}), it is known that the 
superfluid modes become non-radiating, and hence  
${\rm Im}(\omega M) \rightarrow 0$, in the limit 
$\Delta \beta\rightarrow 0$ \citep{andersson-2002-66}. 

While ${\rm Im}(\omega M)$ does not follow any universal law, it is interesting
to note that ${\rm Im}(\omega M)$ can be rescaled in such a way 
that the data from models with the same value of $\bar{\beta}$ lie on the 
same curve. 
In Figure~\ref{fig:sc_fs_IM} we plot ${\rm Im}(\omega M)/\Delta \beta^2$ 
vs $C$ for the models A$i$, B$i$, and C$i$ ($i=2-5$). It is seen clearly that, 
for a given $\bar \beta$ (eg, the models A$i$) and compactness $C$, the 
imaginary parts of the mode frequencies (normalized by the asymmetry parameter 
$\Delta \beta^2$) are rather insensitive to $\Delta \beta$.

The above results suggest that, for the polytropic EOS 
model~(\ref{eq:master_poly}), the imaginary parts of the $f_{\rm s}$-modes are 
given approximately by 
\begin{equation}
\mathrm{Im}(\omega \mathrm{M}) = f \Delta \beta^2 , 
\label{eq:im_fsf1}
\end{equation}
where the value of $f$ depends only on $\bar{\beta}$ and $C$. 
So why is there such a scaling law? In our models we have fixed 
$ \sigma_{\n} = \sigma_{\p} $ and thus we expect our equations to be invariant 
under $ \Delta \beta \rightarrow - \Delta \beta $ and exchange of species 
label $\rm n$ and $\rm p$. As $\mathrm{Im}(\omega \mathrm{M}) = 0$ when 
$\Delta \beta = 0$, we thus expect that the leading dependence of 
$\mathrm{Im}(\omega \mathrm{M})$ on $\Delta \beta$ must be quadratic.
In fact, it can also be shown that the leading dependence is also 
quadratic even if $\sigma_\n \neq \sigma_\p$ \citep{wong_thesis08}. 
We shall present in the Appendix a qualitative argument to explain
why the damping time of the $f_{\rm s}$-mode depends sensitively on various
parameters in the EOS, including $\beta_{\rm n}$, $\beta_{\rm p}$,
$\sigma_{\rm n}$ and $\sigma_{\rm p}$.  Such dependence on the
details of the EOS is definitely an advantage to asteroseismology as
observation of $f_{\rm s}$-mode oscillations is then likely to provide clues
to the coupling between neutron and proton fluids in neutron stars,
which is not yet thoroughly understood.

\subsection{Effects of entrainment} 
\label{sec:num_entrain}

In the above we have demonstrated that the real part of the $f_{\rm s}$-mode
exhibits a universal behavior. 
The value of ${\rm Re}(\omega M)$ depends essentially only on the 
compactness $C$ of the star (for a given value of $l$) and can be 
approximated very well by Eq.~(\ref{eq:newton_fs}). While we have only 
focused on a class of EOS given by Eq.~(\ref{eq:master_poly}), we believe 
that the universal behavior in general holds for decoupled fluids (as defined 
in Sec.~\ref{sec:eos}). However, it is also clear that one must consider the 
couplings between the two fluids in any realistic modeling of superfluid 
neutron stars. 
Here we shall consider the entrainment between the two fluids using 
the master function (\ref{eq:master_lambda01}) in order to study the 
effects of coupling on the oscillation modes. 
It should be noted that the entrainment effect on the ordinary 
$f_{\rm o}$-mode is rather small \citep[see, e.g.,][]{andersson-2002-66}. 
Hence, we shall not consider the $f_{\rm o}$-mode in the study.

We shall focus only on the real part of the $f_{\rm s}$-mode frequency and 
simply refer to it as the oscillation frequency hereafter. 
We use the two EOS models B2 and B5 
(see Table \ref{tab:1}) for the first part $\lambda_0$ of the master function 
(\ref{eq:master_lambda01}). 
It is recalled that the background stellar model is completely 
determined by $\lambda_0$. The entrainment effect is given by the second
part $\lambda_1$ and its strength is governed by the parameter $\epsilon$. 
For a given background EOS model $\lambda_0$ and compactness $C$, we calculate 
the oscillation frequency of the $f_{\rm s}$-mode as a function 
of $\epsilon$. 
The results are summarized in Figures \ref{fig:b2_w2_vs_e} 
and \ref{fig:b5_w2_vs_e}. 

In Figure \ref{fig:b2_w2_vs_e} we plot the squared $f_{\rm s}$-mode 
frequency $\omega (\epsilon)^2$ (normalized by its value when $\epsilon=0$) 
against $\epsilon$ for model B2. 
In the figure, the circle and square data points correspond to the 
compactness $C=0.1$ and $C=0.2$ respectively. 
The corresponding results for model B5 are shown in 
Figure \ref{fig:b5_w2_vs_e}.
As found by \citet{andersson-2002-66}, we see that the mode frequency 
increases with the entrainment parameter $\epsilon$. It is also seen that 
the rate of increase of the mode frequency depends on the compactness of the 
stellar models.

It has been seen in Sec.~\ref{sec:num_decouple} that the oscillation 
frequency of the $f_{\rm s}$-mode essentially does not depend on the 
chosen EOS model for a given compactness. 
Figures \ref{fig:b2_w2_vs_e} and \ref{fig:b5_w2_vs_e} demonstrate 
that the entrainment between the two fluids in general breaks such kind of 
universal behavior. For the case where $\epsilon = 0.1$ and $C=0.1$, the 
ratio $\omega(\epsilon)^2 / \omega(0)^2$ is 1.193 and 1.268 respectively 
for models B2 and B5. Note that the value $\omega(0)$ is the same for both 
models. Nevertheless, we shall show in Sec.~\ref{sec:perturb_poly} 
that the shift in the frequency for a given value of $\epsilon$, namely the 
ratio $\omega(\epsilon)^2 / \omega(0)^2$, can be obtained anlytically from the 
following equation (see Eq.~(\ref{eq:1st_order_rel_final})) 
which involves only the parameters of the background EOS model and the 
compactness: 
\begin{equation}
\begin{split}
{\omega (\epsilon)^2 \over \omega (0)^2}
\approx& 1 + { \epsilon \left[ 1 + { \left( {\tilde \sigma}_{\n} 
\beta_{\n} \right)^{1 / (\beta_{\n} - 1)} \over 
\left( {\tilde \sigma}_{\p} \beta_{\p} \right)^{1 / 
(\beta_{\p} - 1) }} g(0.2)^{{1 \over \beta_{\p} - 1} - {1 \over \beta_{\n} - 1}} \right] \over \left[ 1 + g(0.2) \right] } , 
\end{split}
\label{eq:w2_over_w2f2}
\end{equation}
where the function $g(x)$ is defined in Eq.~(\ref{eq:e_expand}). 
The excellent agreement between the analytic result and numerical data can be 
seen in Figures \ref{fig:b2_w2_vs_e} and \ref{fig:b5_w2_vs_e}. 
The solid and dashed lines in the figures represent 
Eq.~(\ref{eq:w2_over_w2f2}) respectively for $C=0.1$ and $0.2$. 
It is seen clearly that, for the physically reasonable range of $\epsilon$ 
considered by us, Eq.~(\ref{eq:w2_over_w2f2}) agrees with the numerical 
data very well. Significant derivation of the data from 
Eq.~(\ref{eq:w2_over_w2f2}) occurs only for higher values of $\epsilon$.

\section{Superfluid mode for a homogeneous two-fluid star}
\label{sec:fs_mode_derive}

In this section, we shall derive the scaling law (\ref{eq:newton_fs}) for the
real parts of the superfluid $f_{\rm s}$-modes using the Newtonian two-fluid
hydrodynamics equations. Our analysis is based on a model in which the 
densities of the two fluids are constant throughout the interior of the star.
The general set of two-fluid equations in Newtonian gravity 
can be found in \citet{andersson2001dsn}: 
\begin{eqnarray}
0 &=& \partial_t \rho_\n + \partial_i (\rho_\n v_\n^i) , \label{eq:2fluid_1} \\
0 &=& \partial_t \rho_\p + \partial_i (\rho_\p v_\p^i) , \\
0 &=& \partial_t \left[ v_\n^i + \frac{2 \alpha}{\rho_\n} (v_\p^i-v_\n^i) \right] 
+ v_\n^j \partial_j \left[ v_\n^i + \frac{2 \alpha}{\rho_\n} (v_\p^i-v_\n^i) 
\right] \cr
&& \, + \delta^{ij} \partial_j (\Phi + {\tilde \mu}_\n) + \frac{2 \alpha}{\rho_\n} \delta^{ij} \delta_{kl} v_\n^l \partial_j (v_\p^k -v_\n^k) , \\
0 &=& \partial_t \left[ v_\p^i + \frac{2 \alpha}{\rho_\p} (v_\n^i-v_\p^i) \right] 
+ v_\p^j \partial_j \left[ v_\p^i + \frac{2 \alpha}{\rho_\p} (v_\n^i-v_\p^i) 
\right] \cr
&& \, + \delta^{ij} \partial_j (\Phi + {\tilde \mu}_\p) + \frac{2 \alpha}{\rho_\p} \delta^{ij} \delta_{kl} v_\p^l \partial_j (v_\n^k -v_\p^k) , \\
\partial^i \partial_i \Phi &=& 4 \pi G (\rho_\n + \rho_\p) , \label{eq:2fluid_5}
\end{eqnarray}
where $ \rho_\x $, $ v_\x^i $, and $ {\tilde \mu}_\x $ ($ \x $ = $ \n $, $ \p $) are the mass density, velocity, and chemical potential per unit mass of species $ \x $ respectively. $\Phi$ is the gravitational potential. 
$ \alpha $ is a function to describe the entrainment and is defined by 
\begin{equation}
d U (\rho_\n , \rho_\p , \Delta^2 )
= {\tilde \mu}_\n d \rho_\n + {\tilde \mu}_\p d \rho_\p 
+ \alpha d \Delta^2 ,
\end{equation}
where $ U $ is the internal energy density and 
$ \Delta^2 = |{\boldsymbol v}_\n - {\boldsymbol v}_\p |^2 $. Note that the relativistic coefficient $ \mathcal{A} $ 
defined in Eq.~(\ref{eq:abc_coeff}) is proportional to $ \alpha $ in the 
Newtonian limit \citep{andersson2001dsn}. 

For a non-rotating static background ($ v_\x^i = 0 $), the equilibrium 
equations are given by
\begin{eqnarray}
 \partial_i (\Phi + {\tilde \mu}_\x) &=& 0 ,
\label{eq:bkg1}\\ 
 \partial^i \partial_i \Phi &=& 4 \pi G (\rho_\n + \rho_\p) .
\end{eqnarray}
In particular, we shall assume that the background star is homogeneous
in which $\rho_\n$ and $\rho_\p$ are constants. The solutions for the 
background equations are then given by
\begin{eqnarray}
\Phi &=& 
\left\{ 
\begin{array}{lll}
- \frac{4 \pi G }{6} (\rho_{\n} + \rho_{\p}) \left( 3 R^2 - r^2 \right), \quad & r < R \\
\\
- \frac{4 \pi G }{3 r}  (\rho_{\n} + \rho_{\p}) R^3 , \quad & r \geq R
\end{array}
\right.  \\
\\ 
{\tilde \mu}_{\x} &=& 
\left\{ 
\begin{array}{lll}
\frac{4 \pi}{6} G (\rho_{\n} + \rho_{\p}) \left( R^2 - r^2 \right), \quad & r < R \\
\\
0, \quad & r \geq R
\end{array}
\right. 
\label{eq:mu_sol}
\end{eqnarray}

The linearized versions of Eqs.~(\ref{eq:2fluid_1})-(\ref{eq:2fluid_5}) 
have been studied previously 
\citep[e.g.,][]{lindblom1994osn,andersson2001dsn,prix2002aon}.
In particular, the linearized hydrodynamics equations for a non-rotating 
static background are 
\begin{equation}
\partial_{t}^2 \left[ \xi_{\n}^{i} + \frac{2 \alpha}{\rho_{\n}}(\xi_{\p}^{i} - \xi_{\n}^{i}) \right]  
= -\partial_{i} (\delta \Phi + \delta {\tilde \mu}_{\n}) ,
\label{eq:en_pm1}
\end{equation}
\begin{equation}
\partial_{t}^2 \left[ \xi_{\p}^{i} + \frac{2 \alpha}{\rho_{\p}}(\xi_{\n}^{i} - \xi_{\p}^{i}) \right]  
= -\partial_{i} (\delta \Phi + \delta {\tilde \mu}_{\p}) ,
\label{eq:en_pm2}
\end{equation}
\begin{equation}
\delta \rho_{\n} + \partial_{i} \left( \rho_{\n} \xi_{\n}^{i} \right) = 0 ,
\label{eq:en_pd1}
\end{equation}
\begin{equation}
\delta \rho_{\p} + \partial_{i} \left( \rho_{\p} \xi_{\p}^{i} \right) = 0 ,
\label{eq:en_pd2}
\end{equation}
\begin{equation}
\partial_{j} \partial_{j} \delta \Phi = 4 \pi G 
\left( \delta \rho_{\n} + \delta \rho_{\p}  \right) , 
\label{eq:en_pg1}
\end{equation}
where $ \xi^{i}_{\x} $ is the Lagrangian displacements for species $ \x $ and 
$\delta$ is used to denote Eulerian perturbations.
For simplicity, we shall set $\alpha=0$ in the following analysis. 
The effect of entrainment will be studied in Sec.~\ref{sec:pert}.

For our homogeneous background model, together with the assumption that the 
two fluids are decoupled, it can be shown from the linearized hydrodynamics
equations that the superfluid modes are governed by the 
following equations \citep{andersson2001dsn}:
\begin{equation}
{\partial^2 \bdxi_{-} \over \partial t^2} + {\bf \nabla} \delta \beta 
= 0 , 
\label{eq:xi_dt}
\end{equation}
\begin{equation}
{\bf \nabla} \cdot \bdxi_{-} = 0 ,
\label{eq:xi_dig} 
\end{equation}
where
$\partial \bdxi_{-} / \partial t \equiv \delta {\boldsymbol v}_\p 
- \delta {\boldsymbol v}_\n$ and $\delta \beta \equiv \delta {\tilde \mu}_\p 
- \delta {\tilde \mu}_\n$. A nonzero relative velocity 
$\bdxi_{-}$ is a characteristic of the superfluid modes. 
For our simplified stellar model, we note that the equations for the 
superfluid modes are completely decoupled from those for the ordinary-fluid 
modes \citep{andersson2001dsn,prix2002aon}.
Furthermore, the superfluid modes are completely decoupled from the 
perturbation of the gravitational potential $\delta \Phi$. The
counter-motion between the two fluids implies that the total 
density variation $\delta \rho = \delta \rho_{\rm n} + \delta \rho_{\rm p}$,
and hence $\delta \Phi$, nearly vanish.   
Combining Eqs.~(\ref{eq:xi_dt}) and (\ref{eq:xi_dig}), we obtain
\begin{equation}
\nabla^2 \delta \beta = 0 . 
\label{eq:smd4}
\end{equation}
In solving Eq.~(\ref{eq:smd4}), boundary conditions should be imposed at the
center and the stellar surface: (i) the solution should be regular at $r=0$; 
(ii) the Lagrangian variation of $\beta \equiv {\tilde \mu}_\p
-{\tilde \mu}_{\n}$ should vanish at $r=R$. 

The background equation (\ref{eq:bkg1}) implies that 
${\tilde \mu}^\prime_\n = {\tilde \mu}^\prime_\p$,
where a ``prime'' denotes radial derivative. The boundary condition at the 
surface can thus be written as
\begin{equation}
\delta \beta + (\xi_\p^r - \xi_\n^r) {\tilde \mu}_{\n}' = 0 ,
\label{eq:bc_R}
\end{equation}
where $\xi_\x^r$ is the radial component of the Lagrangian displacement
of the species $\x$. Eqs.~(\ref{eq:xi_dt}) and (\ref{eq:bc_R}) imply that 
$\delta \beta$ satisfies the following equation at the surface:
\begin{equation}
{\partial^2 \delta \beta \over \partial t^2} - 
{\tilde \mu}^\prime_\n \delta \beta^\prime = 0 .
\label{eq:bc_R_dt}
\end{equation}
Now let us decompose $\delta \beta$ in the form $\delta\beta = 
\delta\beta_{lm} (r) Y_{lm}(\theta,\phi) e^{i \omega t} $. 
Eq.~(\ref{eq:smd4}) and the regularity condition at $r=0$ imply that 
\begin{equation}
\delta \beta_{lm} (r) = A r^l ,
\end{equation}
where $A$ is some constant. Using the radial background profile for 
${\tilde \mu}_\n$ as given in Eq.~(\ref{eq:mu_sol}), the boundary condition
(\ref{eq:bc_R_dt}) implies that the oscillation frequency $\omega$ 
of the superfluid modes is given by 
\begin{equation}
\omega^2 = l {G M \over R^3} . 
\label{eq:newton_fs_derive}
\end{equation} 
This is the scaling law (\ref{eq:newton_fs}) presented in 
Sec.~\ref{sec:num_decouple}, except that we have 
restored the gravitational constant $G$.
As we have seen in Sec.~\ref{sec:num_decouple}, Eq.~(\ref{eq:newton_fs}) 
can describe the data very well as long as $C \lesssim 0.2$.
Relativistic effects become important for higher compactness. 
As can be seen from Figures~\ref{fig:fs_RE_B} and \ref{fig:fs_RE_C}, 
the derivation of the numerical data from Eq.~(\ref{eq:newton_fs_derive}) 
becomes noticeable for $C \gtrsim 0.2$. 
To further illustrate the relativistic corrections, we plot in 
Figure~\ref{fig:modelC5_rel_diff} the relative difference between the 
numerical data and the analytic result~(\ref{eq:newton_fs_derive}) for 
the $f_{\rm s}$-mode frequency of model C5. We see that the relative 
difference increases linearly with the compactness of the star. 
The relativistic correction rises up to about 8\% for the maximum mass 
configuration. 

It is worthy to point out that Eq.~(\ref{eq:newton_fs_derive}) is 
in fact also the result for the $f$-mode of an incompressible single-fluid 
star in the Cowling approximation (i.e., assuming $\delta \Phi = 0 $). 
For such a single-fluid model, it can be shown that the perturbed one-fluid 
equations governing the $f$-mode reduce to the same equation (\ref{eq:smd4}) 
with $\delta \beta$ being replaced by the perturbed enthalpy 
$\delta h = \delta P / \rho$:
\begin{equation}
\nabla^2 \delta h = 0 . 
\label{eq:perturb_h}
\end{equation} 
The equivalence between the two governing equations (\ref{eq:smd4}) and 
(\ref{eq:perturb_h}) 
comes from the fact that the counter-motion of a superfluid mode leads to a 
nearly vanishing $\delta \Phi$ as discussed above. 
However, it should be noted that Eq.~(\ref{eq:newton_fs_derive}) does not 
describe the numerical data of ordinary $f$-modes as good as it 
does for the superfluid $f$-modes. This suggests that incompressiblity and 
the Cowling approximation are not good approximations for the 
ordinary $f$-modes 
\citep[see][for the analysis of the ordinary $f$-mode with the 
compressibility taken into account]{andersson-2008}.

Finally, if we also assume that the entrainment term $\alpha$ is 
a constant, Eq.~(\ref{eq:newton_fs_derive}) can be easily generalized to 
include the effect of entrainment and becomes
\begin{equation}
\omega^2 = l { G M \over R^3} \left( 1 + \epsilon 
{ \rho_{\rm n} \over \rho_{\rm p} } \right) ,
\label{eq:newton_fs_with_epsilon}
\end{equation}
where the two parameters $\alpha$ and $\epsilon$ are related by 
$\alpha = \epsilon \rho_{\rm n} \rho_{\rm p} / 
[ 2 ( \epsilon \rho_{\rm n} + ( 1 + \epsilon) \rho_{\rm p} ) ] $.
At first sight, one might expect that this generalized result 
could be used directly to compare with the numerical data. However, it 
should be noticed that Eq.~(\ref{eq:newton_fs_with_epsilon}) involves 
the mass densities of the two fluid components. This makes the comparison
not practical since the numerical data are obtained from stellar models 
where $\rho_{\rm n}$ and $\rho_{\rm p}$ are varying throughout the stars. 
In the next section, we shall use a variational principle to study the 
effect of entrainment perturbatively.

\section{Pertubative Analysis of the entrainment}
\label{sec:pert}

\subsection{General integral formula}

In this section we shall study the effect of entrainment on the superfluid 
mode frequency for a two-fluid star without rotation. Our approach is to treat
the entrainment as a perturbation and employ a variational principle to 
calculate the first-order shift in the mode frequency. We have generalized
the variational principle for polar oscillation modes of ordinary-fluid 
relativistic stars developed by \citet{detweiler_1973} to the case of 
two-fluid stars.  The derivation of the general relativistic variational 
principle is somewhat lengthy and we shall not present the details in 
this paper \citep[see][]{wong_thesis08}.
However, in order to illustrate the basic idea, 
we shall first derive the corresponding variational principle in the 
Newtonian framework and then simply quote the relativistic result. 

The relevant equations for the Newtonian analysis are given by 
Eqs.~(\ref{eq:en_pm1})-(\ref{eq:en_pg1}). 
We shall treat this system of equations as an 
eigenvalue system with the operator containing the entrainment term
$\alpha$ as the perturbing Hamiltonian. 
Consider a normal-mode solution with the time dependence  
of the form $\delta f( \bdr, t) \equiv \delta f(\bdr) 
e^{i\omega t}$, where $\delta f$ is a perturbed quantity. 
Eqs.~(\ref{eq:en_pm1}) and (\ref{eq:en_pm2}) then become 
\begin{eqnarray}
\omega^2 \left[\xi_{\n}^{i} + \frac{2 \alpha}{\rho_{\n}}(\xi_{\p}^{i} - \xi_{\n}^{i}) \right]  &=& \partial_{i} (\delta \Phi + \delta {\tilde \mu}_{\n}) ,
\label{eq:en_pm1_omega} \\
&& \cr
\omega^2 \left[ \xi_{\p}^{i} + \frac{2 \alpha}{\rho_{\p}}(\xi_{\n}^{i} - \xi_{\p}^{i}) \right]  &=& \partial_{i} (\delta \Phi + \delta {\tilde \mu}_{\p}) .
\label{eq:en_pm2_omega} 
\end{eqnarray}
We want to formally express the above equations (to first-order of smallness 
in $\alpha$) in the form 
\begin{equation}
\left( {\hat H}_{(0)} + {\hat H}_{(1)} \right) {\bf Y} = 
\omega^2 {\bf Y},
\end{equation}
where  $\hat{H}_{(i)}$ is an operator $i$-th order in $\alpha$ and 
${\bf Y} \equiv (\bdxi_\n , \bdxi_\p )$ is an abstract 
eigenvector with the two Lagrangian displacement vectors as its components.
The system is solved with the ``constraint'' equations 
(\ref{eq:en_pd1})-(\ref{eq:en_pg1}) which do not contain time derivatives. 

First we define the ``unperturbed Hamiltonian'' ${\hat H}_{(0)}$ by 
\begin{equation}
\hat{H}_{(0)} \left( \begin{matrix} 
\bdxi_{\n} \\
\bdxi_{\p} \end{matrix} 
\right) = 
\left( \begin{matrix}
\partial_{i} (\delta \Phi + \delta {\tilde \mu}_{\n} ) \\
\partial_{i} (\delta \Phi + \delta {\tilde \mu}_{\p} )
\end{matrix}
\right) .
\end{equation}
In order to apply standard perturbation theory as in quantum mechanics, we 
first need to define an inner product such that ${\hat H}_{(0)}$ is symmetric.
To this end, we define the inner product
\begin{equation}
(\bdeta_{\n}, \bdeta_{\p}) \cdot (\bdxi_{\n}, \bdxi_{\p}) = \int dV (\rho_{\n} \bdeta_{\n} \cdot \bdxi_{\n} + \rho_{\p} \bdeta_{\p} \cdot \bdxi_{\p}) .
\end{equation}
We shall show that, under this inner product, ${\hat H}_{(0)}$ is symmetric.
In other words,  
$ (\bdeta_{\n}, \bdeta_{\p}) \cdot {\hat H}_{(0)} 
\left( \begin{matrix} \bdxi_{\n} \\ \bdxi_{\p} \end{matrix} \right) $ 
is symmetic in $ (\bdeta_{\n}, \bdeta_{\p}) $ and 
$ (\bdxi_{\n}, \bdxi_{\p}) $:
\begin{equation}
\begin{split}
\quad & (\bdeta_{\n}, \bdeta_{\p}) \cdot \hat{H}_{(0)} \left( \begin{matrix} 
\bdxi_{\n} \\ \bdxi_{\p} 
\end{matrix} \right) \\
 =&
\int dV \left[ (\rho_{\n} \eta_{\n}^i) \partial_i (\delta \Phi + 
\delta {\tilde \mu}_{\n} ) + 
(\rho_{\p} \eta_{\p}^i) \partial_i (\delta \Phi + \delta {\tilde \mu}_{\p})
\right] 
\\
 =& \int dS \, \hat{r}_j \left[ \rho_{\n} \eta_{\n}^j 
(\delta \Phi + \delta {\tilde \mu}_{\n}) + \rho_{\p} \eta_{\p}^j 
(\delta \Phi +  \delta {\tilde \mu}_{\p} ) \right] \\
& + \int dV \left[ {\hat \delta}\rho_{\n} 
(\delta \Phi + \delta {\tilde \mu}_{\n} ) + 
{\hat \delta}\rho_{\p} (\delta \Phi +  \delta {\tilde \mu}_{\p} ) \right] ,
\end{split}
\label{eq:int}
\end{equation}
where the first term is a surface integral which is carried out on the two 
sphere $ r = R $ and $ \hat{\boldsymbol r} $ is the outward unit normal to the 
sphere. 
To arrive at the second row, we have performed an integration by parts and 
used Eqs.~(\ref{eq:en_pd1}) and (\ref{eq:en_pd2}). 
$\delta f$ and $ {\hat \delta} f $ refer to the Eulerian 
variation of the quantity $f$ associated to the displacement 
$\bdxi_{\x}$ and $\bdeta_{\x}$ respectively.  

Note that the surface integral in Eq.~(\ref{eq:int}) vanishes since the 
densities tend to zeros at the surface for our stellar models. 
The remaining volume integral can be shown to be symmetric in  
$ (\bdeta_{\n}, \bdeta_{\p}) $ and 
$ (\bdxi_{\n}, \bdxi_{\p}) $. To show this, we first note that  
\begin{equation}
\begin{split}
{\tilde \mu}_{\n} =& \, \frac{\partial U (\rho_{\n}, \rho_{\p}, \Delta^2)}{\partial \rho_{\n}} ,
\end{split}
\end{equation}
and thus
\begin{equation}
\begin{split}
\delta {\tilde \mu}_{\n} =& \, \frac{\partial^2 U}{\partial \rho_{\n}^2} \delta \rho_{\n} + \frac{\partial U}{\partial \rho_{\n} \partial \rho_{\p}} \delta \rho_{\p} .
\end{split}
\end{equation}
It should be mentioned that a term associated with $ d \Delta^2 $ vanishes because in the background $ \Delta = 0 $. This implies that 
\begin{equation}
\begin{split}
\hat{\delta}\rho_{\n} \delta {\tilde \mu}_{\n} + \hat{\delta}\rho_{\p} 
\delta {\tilde \mu}_{\p} 
=& \, \hat{\delta}\rho_{\n} \delta \rho_{\n} \frac{\partial^2 U}{\partial \rho_{\n}^2} + \hat{\delta}\rho_{\p} \delta \rho_{\p} \frac{\partial^2 U}{\partial \rho_{\p}^2} 
\\
& \, + (\hat{\delta}\rho_{\p} \delta \rho_{\n} + \hat{\delta}\rho_{\n} \delta \rho_{\p}) \frac{\partial U}{\partial \rho_{\n} \partial \rho_{\p}} ,
\end{split}
\end{equation}
which is symmetric in $ (\hat{\delta}\rho_{\n}, \hat{\delta}\rho_{\p}) $ and $ (\delta \rho_{\n}, \delta \rho_{\p}) $. 
It can also be shown easily that 
$ \int dV (\hat{\delta}\rho_{\n} + \hat{\delta}\rho_{\p}) 
\delta \Phi $ is symmetric. 
Thus the volume integral in Eq.(\ref{eq:int}), and hence 
${\hat H}_{(0)}$, is symmetric.

We can now apply standard perturbation theory as in quantum mechanics 
with the perturbing potential ${\hat H}_{(1)}$ defined by 
\begin{equation}
\hat{H}_{(1)} = 2 \alpha \omega^2_{(0)}
\left( \begin{matrix}
1 / \rho_{\n} & - 1 / \rho_{\n} \\
- 1 / \rho_{\p}  & 1 / \rho_{\p} 
\end{matrix}
\right) ,
\end{equation}
where $ \omega^2_{(0)} $ denotes the squared frequency of the mode in the 
absence of entrainment (i.e. $ \alpha = 0 $). The first-order shift in the 
squared frequency is given by 
\begin{equation}
\begin{split}
\omega^2_{(1)} =& \, { (\bdxi_{\n}^{\ (0)}, \bdxi_{\p}^{\ (0)} ) 
\cdot \hat{H}_{(1)} \left( \begin{matrix}
\bdxi_{\n}^{\ (0)} \\
\bdxi_{\p}^{\ (0)}
\end{matrix}
\right) \over \left[ ( \bdxi_{\n}^{\ (0)}, \bdxi_{\p}^{\ (0)} ) 
\cdot ( \bdxi_{\n}^{\ (0)}, \bdxi_{\p}^{\ (0)} ) \right] }\\
=& \, \frac{2 \omega^2_{(0)} \int dV \left
(\alpha \left| \bdxi_{\n}^{\ (0)} 
- \bdxi_{\p}^{\ (0)} \right|^2 \right)}{\int dV \left( \rho_{\n} 
\bdxi_{\n}^{\ (0)} 
\cdot \bdxi_{\n}^{\ (0)} + \rho_{\p} \bdxi_{\p}^{\ (0)} \cdot 
\bdxi_{\p}^{\ (0)} \right)} .
\end{split}
\label{eq:1st_order}
\end{equation} 
where $\bdxi_{\n}^{\ (0)}$ and $\bdxi_{\p}^{\ (0)}$ are the 
Lagrangian displacement vectors associated to the ``zeroth-order'' mode 
solution with squared frequency $\omega^2_{(0)}$. 

For a given static background of two-fluid stellar model, the effect of 
a small entrainment on the oscillation mode frequency is to change the 
squared frequency $\omega_{(0)}^2$ to 
$\omega^2 = \omega_{(0)}^2 + \omega_{(1)}^2$. 
It can also be seen qualitatively from Eq.~(\ref{eq:1st_order}) that 
the ordinary-fluid modes in general do not depend on entrainment 
\citep[as has been discussed in][]{andersson2001dsn,andersson-2002-66,prix2002aon}.
The first-order shift depends on the relative displacement
$| \bdxi_{\n}^{\ (0)} - \bdxi_{\p}^{\ (0)} |$, which tends to zero
for ordinary-fluid modes.

Eq.~(\ref{eq:1st_order}) can readily be used to compute the change in the 
mode frequency once the entrainment function $\alpha$ and the zeroth-order 
solution $(\bdxi_\n^{(0)} , \bdxi_\p^{(0)})$ are given. However, we cannot 
employ the equation directly since it is derived in the Newtonian framework, 
while our zeroth-order solution $(\bdxi_\n^{(0)} , \bdxi_\p^{(0)})$ 
are computed using a relativistic numerical code. 
Nevertheless, with the derivation of 
Eq.~(\ref{eq:1st_order}) as an illustration of the basic idea involved, we 
have derived the corresponding result based on a two-fluid formalism 
extension of the work of \citet{detweiler_1973}. The derivation of the 
relativistic case is somewhat tedious and lengthy \citep{wong_thesis08}.
Here we shall only present the final result of the first-order shift in 
$\omega^2$:
\begin{equation}
\omega^2_{(1)} = 
 -  \frac{\omega_{(0)}^2 \int dV_3 e^{-\nu /2 }\left(\mathcal{A} n p \: \left| \bdxi_{\n}^{(0)} - \bdxi_{\p}^{(0)} \right|^2 \right)}
{\int dV_3 e^{- \nu / 2 }
\left[ {\partial \lambda_0 \over \partial n} n 
( \bdxi_{\n}^{(0)} \cdot \bdxi_{\n}^{(0)} )
+ {\partial \lambda_0 \over \partial p} p 
( \bdxi_{\p}^{(0)} \cdot \bdxi_{\p}^{(0)} ) \right]} .
\label{eq:1st_order_rel}
\end{equation}
where $d V_3 =  e^{\lambda / 2} r^2 \sin \theta dr d\theta d\phi$ is the 
proper volume element for the spatial 3-geometry of the background 
spacetime metric 
\begin{equation}
ds^2 = -e^{\nu(r)} dt^2 + e^{\lambda(r)} dr^2 + r^2 \left( d\theta^2 + \sin^2\theta d\phi^2 \right) .
\end{equation}
The 3-vectors $\bdxi_\n^{(0)}$ and $\bdxi_\p^{(0)}$ are still defined as 
the Lagrangian displacement vectors associated to the zeroth-order solution
as before. But the scalar product between any two 3-vectors $\boldsymbol U$
and $\boldsymbol V$ is now given by 
\begin{equation}
{\boldsymbol U} \cdot {\boldsymbol V} 
= e^{\lambda} U^r V^r + r^2 U^{\theta} V^{\theta} + r^2 \sin^2 \theta 
U^{\phi} V^{\phi} , 
\end{equation}
where the $U^i$'s and $V^i$'s are the coordinate components of the vectors. 
The entrainment is now described by the function $\cal A$ as discussed in 
Sec.~\ref{sec:formal}.
It can also be shown easily that Eq.~(\ref{eq:1st_order_rel}) reduces to 
Eq.~(\ref{eq:1st_order}) in the Newtonian limit.

\subsection{Decoupled polytropes}
\label{sec:perturb_poly}

Eq.~(\ref{eq:1st_order_rel}) is a general result and could be used to obtain 
the frequency shift due to a small entrainment once the zeroth-order mode 
solution (i.e., without the entrainment) is given. 
In deriving Eq.~(\ref{eq:1st_order_rel}) we have not made any assumption about 
the master function $\Lambda$. 
In this subsection, we focus on the special case where the master function 
$\Lambda$ is given by Eq.~(\ref{eq:master_lambda01}). We shall show that 
the integral formula (\ref{eq:1st_order_rel}) can be approximated very well 
by an algebraic relation which involves only the parameters of the 
background EOS (i.e., the part $\lambda_0$ in Eq.~(\ref{eq:master_lambda01})), 
the entrainment parameter $\epsilon$, and the compactness of the star $C$.

To obtain an approximation to Eq.~(\ref{eq:1st_order_rel}), we first note 
that the chemical potentials of the neutrons and the (conglomerate)
protons evaluated on the static background are respectively 
\citep{comer-1999-60}
\begin{equation}
\mu = {\cal B} n + {\cal A} p = \mu_\infty e^{- \nu /2} , \ \ \ 
\chi = {\cal C } p + {\cal A} = \chi_\infty e^{- \nu /2} ,  
\end{equation}
where $\mu_\infty$ and $\chi_\infty$ are constants. The condition of 
chemical equilibrium, with Eq.~(\ref{eq:thm_coeff}) for the thermodynamic 
coefficients, implies that 
 \begin{equation}
{\partial \lambda_0 \over \partial n} = {\partial \lambda_0 \over \partial p} = \mu_{\infty} e^{-{\nu \over 2}} .
\end{equation}
Using this relation, Eq.~(\ref{eq:1st_order_rel}) can now be written as 
\begin{equation}
\begin{split}
\omega^2_{(1)} 
=& - {1 \over \mu_{\infty}} \frac{\omega_{(0)}^2 \int 
dV_3 e^{- \nu / 2 } \left( \mathcal{A} n p \: 
\left| \bdD \right|^2 \right)}
{\int dV_3 e^{-\nu}\left[ \left(n+p\right) \left| \bdU \right|^2 + 
{n p \over n+p} \left| \bdD \right|^2 \right] } ,
\end{split}
\label{eq:1st_order_relf2}
\end{equation}
where the 3-vectors $\bdU$ and $\bdD$ are defined by 
\begin{equation}
\bdU \equiv {n  \bdxi_{\n}^{(0)} + p  \bdxi_{\p}^{(0)} \over n+p}, 
\quad  \bdD \equiv \bdxi_{\n}^{(0)} - \bdxi_{\p}^{(0)} .
\end{equation}

We note that for the superfluid modes where the two fluids are 
dominated by counter-moving motion (i.e., when $|\bdD|$ is large), 
Eq.~(\ref{eq:1st_order_relf2}) can be approximated by 
\begin{equation}
\begin{split}
\omega^2_{(1)} 
\approx& - {1 \over \mu_{\infty}} \frac{\omega_{(0)}^2 \int 
dV_3 e^{-\nu /2 }(\mathcal{A} n p \: \left| \bdD \right|^2)}{\int dV_3 e^{-\nu}\left({n p \over n+p} \left| \bdD \right|^2 \right)} .
\end{split}
\label{eq:1st_order_relf3}
\end{equation}
To show that this is in general a good approximation, we define the ratio 
\begin{eqnarray}
\eta \equiv
{\int dV_3 e^{-\nu}\left[ \left(n+p\right) \left| \bdU \right|^2 \right] \over
\int dV_3 e^{-\nu}\left( {n p \over n+p} \left| \bdD \right|^2 \right)} ,
\label{eq:ratio_ud}
\end{eqnarray}
and present its value for the $f_{\rm s}$-modes of a few typical stellar 
models in Table~\ref{tab:U}. It is seen that $\eta$ is in general much 
smaller than unity. Hence, it is a good approximation to neglect the 
first integral in the denominator of Eq.~(\ref{eq:1st_order_relf2}).

In the relativistic perturbative formalism \citep{comer-1999-60}, 
the Lagrangian displacements are decomposed into 
\begin{eqnarray}
   \xi_{\rm x}^r &=& e^{-\lambda/2} r^{l - 1}  W_{\rm x}(r) P_l(\cos \theta) , 
\nonumber\\                   
   \xi_{\rm x}^{\theta} &=& - r^{l - 2} V_{\rm x}(r) 
                {\partial \over \partial \theta} P_l(\cos \theta) ,
\label{eq:lgr_dpm}
\end{eqnarray}
where $P_l(x)$ is the Lengendre polynomial. 
Hence, Eq.~(\ref{eq:1st_order_relf3}) can be put into the following form
\begin{equation}
\begin{split}
{\omega^2_{(1)} \over \omega_{(0)}^2}
\approx& - {1 \over \mu_{\infty}}  \frac{\int dr r^{2 l} e^{\frac{\lambda-\nu}{2}} \mathcal{A} n p {\tilde D}(r)}{\int dr r^{2l} e^{\frac{\lambda}{2}-\nu} \left({n p \over n+p} {\tilde D}(r) \right)} ,
\end{split}
\label{eq:1st_order_relf4}
\end{equation}
where
\begin{eqnarray}
{\tilde D}(r) &\equiv& e^{-\lambda(r)} \left[ W_{\n}(r)-W_{\p}(r)\right]^2 \nonumber\\
&&+ l(l+1) \left[ V_{\n}(r) - V_{\p}(r) \right]^2 .
\end{eqnarray}
We also observe that the two functions (which are the two integrands 
in Eq.~(\ref{eq:1st_order_relf4})) 
\begin{eqnarray}
f_1(r)&\equiv& r^{2l} e^{\frac{\lambda-\nu}{2}} \mathcal{A} n p {\tilde D}(r) ,
\nonumber\\
f_2(r) &\equiv& r^{2l} e^{\frac{\lambda}{2}-\nu} {n p \over n+p} {\tilde D}(r) ,
\label{eq:f1f2}
\end{eqnarray}
have very similar shapes and both reach a maximum value at the position
$r \approx 0.8 R$ (see Figure~\ref{fig:B2B5term}). 
We thus further approximate Eq.~(\ref{eq:1st_order_relf4}) by 
\begin{eqnarray}
{\omega^2_{(1)} \over \omega_{(0)}^2}
&\approx& {1 \over \mu_{\infty}}{f_1(0.8 R) \over f_2(0.8 R)} \nonumber\\
&\approx& \left[ {1 \over \mu_{\infty} e^{- \nu /2} } \left( n+p \right) \mathcal{A} \right]_{r=0.8R} .
\label{eq:1st_order_relf5}
\end{eqnarray}
For the decoupled polytropic model $\lambda_0$ 
(see Eq.~(\ref{eq:master_poly})) we use for the background, we have 
\begin{eqnarray}
\mu &=& \mu_{\infty} e^{-{\nu \over 2}} = m + \sigma_{\n} \beta_{\n} n^{\beta_{\n}-1} ,
\\
\chi &=& \mu_{\infty} e^{-{\nu \over 2}} = m + \sigma_{\p} \beta_{\p} p^{\beta_{\p}-1} .
\end{eqnarray}
Since $n(R) = p(R) = 0$ and $e^{\nu (R)} = \left(1 - 2 C\right)$, we deduce 
that 
\begin{eqnarray}
{\mu_{\infty} \over m} = \sqrt{1 - 2 C } .
\end{eqnarray}
Furthermore, it can be shown that 
\begin{eqnarray}
n(r) &=& \left[{1 \over \tilde{\sigma}_{\n} \beta_{\n}} \left( {\mu_{\infty} \over m} e^{- \nu / 2 } - 1 \right) \right]^{1 / (\beta_{\n} - 1 )} ,
\\
p(r) &=& \left[{1 \over \tilde{\sigma}_{\p} \beta_{\p}} \left( {\mu_{\infty} \over m} e^{-{\nu \over 2}} - 1 \right) \right]^{1 / (\beta_{\p} - 1 ) } ,
\end{eqnarray}
where $\tilde{\sigma}_{\x} = \sigma_{\x} / m$, and
\begin{eqnarray}
{\mu_{\infty} \over m}e^{- \nu(r) / 2 } &\approx & 1 + \frac{C}
{\left(1 - 2C \right)} x + \frac{ C \left(2 - C \right)}{2 \left(1 
- 2 C \right)^2} x^2 \nonumber\\
&\equiv& 1 + g(x) , 
\label{eq:e_expand}
\end{eqnarray}
where $ x \equiv 1- r / R $. 

Using the above relations, we obtain the final expression
\begin{equation}
{\omega^2_{(1)} \over \omega_{(0)}^2} \approx
 { \epsilon \left[ 1 + { \left( \tilde{\sigma}_{\n} \beta_{\n} 
\right)^{1 / (\beta_{\n} - 1)} 
\over \left( \tilde{\sigma}_{\p} \beta_{\p} \right)^{1 / (\beta_{\p} - 1)}} 
g(0.2)^{{1 \over \beta_{\p} - 1} - {1 \over \beta_{\n} - 1}} \right] 
\over \left( 1 + g(0.2) \right)} .
\label{eq:1st_order_rel_final}
\end{equation}
It should be noted that this final formula only depends on the parameters 
of the master function and the compactness of the model. In particular, in
contrast to the ``exact'' integral formula (\ref{eq:1st_order_rel}),
it does not depend on the zeroth-order mode functions 
$(\bdxi_\n^{(0)} ,\bdxi_\p^{(0)})$ which needed to be determined numerically.

In Sec.~\ref{sec:num_entrain} we have seen how well 
Eq.~(\ref{eq:1st_order_rel_final}) agrees to the numerical data. 
Here in Figure~\ref{fig:b2_w2_vs_e2} we demonstrate the good agreement 
again by plotting 
the numerical data, the ``exact'' integral formula 
(\ref{eq:1st_order_rel}), and the approximated formula 
(\ref{eq:1st_order_rel_final}) for the $f_{\rm s}$-modes of EOS model B2
at compactness $C=0.15$. In the figure, we plot the (normalized)
squared $f_{\rm s}$-mode frequency $\omega^2$ against the entrainment 
parameter $\epsilon$ (see Eq.~(\ref{eq:master_lambda1})). 
The circle data points are obtained directly from the relativistic numerical 
code. The solid and dashed lines are obtained respectively 
from Eqs.~(\ref{eq:1st_order_rel}) and (\ref{eq:1st_order_rel_final}). 
It is seen clearly the excellent agreement among the three results for the 
physical range of $\epsilon$ we consider.

\section{Conclusions}
\label{sec:conclude}

In this work we have studied whether the $f_{\rm o}$- and 
$f_{\rm s}$-modes of superfluid neutron stars exhibit any kind of 
universal scaling laws as have been seen in the $f$-mode of ordinary-fluid 
neutron star models. 
We first focus on the simplified case of two decoupled fluids, each with a 
polytropic EOS. We vary the polytropic indices to mimic the 
effects of different EOSs. Our numerical 
results show that the $f_{\rm o}$-mode, where the two fluids move in 
``lock-step'', obeys the same universal scaling laws as the $f$-mode of 
ordinary fluid stars. 

On the other hand, we find that the oscillation frequency of the 
$f_{\rm s}$-mode, which corresponds essentially to counter motion between 
the two fluids, obeys a different scaling law (Eq.~(\ref{eq:newton_fs})). 
We have also derived the scaling law analytically based on 
a homogeneous two-fluid stellar model in Newtonian gravity. 
However, the damping time of the $f_{\rm s}$-mode in general does not exhibit 
any kind of universality. 
While we have only used a generalized polytropic EOS in our 
study, we believe that our conclusion holds in general for superfluid neutron 
star models in which the two fluids exist throughout the whole star and are 
decoupled in the sense that the master function $\Lambda$ (ie, the EOS) can 
be decomposed into two contributions corresponding to each fluid, i.e., 
$\Lambda(n^2, p^2) = \Lambda_{\n} (n^2) + \Lambda_{\p} (p^2)$. 

The inclusion of a coupling term in the master function will in general 
break the universal behavior. To illustrate the effect of coupling, we have 
studied the entrainment between the two fluids using a parameterized 
entrainment model. 
We show numerically how the $f_{\rm s}$-mode frequency increases with the 
strength of the entrainment. Furthermore, based on a relativistic variational 
principle, we have carried out a perturbative analysis and have derived an 
expression for the first-order shift of the frequency due to the entrainment.  
If the superfluid $f_{\rm s}$-modes could be detected by future gravitational
wave detectors, then the derivation of the observed mode frequencies from the
universal scaling curve for the decoupled fluids could then be a useful
probe to the coupling effects between the neutron superfluid and normal fluids
inside neutron stars. 
In summary, our main results (Eqs.~(\ref{eq:newton_fs}) and 
(\ref{eq:w2_over_w2f2})) can be used to obtain a good approximation to the 
oscillation frequency of the $f_{\rm s}$-modes for 
the generalized polytropic EOS and entrainment models that have been used
extensively to study superfluid neutron stars
\citep{comer-1999-60,andersson2001srg,andersson-2002-66,prix2002aon,
yoshida03}.

While we have not yet detected the gravitational 
waves emitted by neutron stars, it is worthy to mention that the first 
gravitational-wave search sensitive to the $f$-modes has recently 
been carried out by the LIGO detectors and interesting upper limits on the 
wave strain (within the predicted range of some theoretical models) have 
also been placed \citep{abbott-2008sgr}.
In view of the fact that the advanced LIGO detectors will have more than a 
factor of 10 improvement on the sensitivity of the wave strain, 
gravitational-wave astroseismology may soon become a reality.

\begin{acknowledgments}
This work is supported in part by the Hong Kong Research Grants Council (Grant
No: 401807) and the direct grant (Project ID: 2060330) from the Chinese 
University of Hong Kong.

\end{acknowledgments}


\appendix

\section{Gravitational-wave emission of superfluid modes}
\label{sec:appendix}

To gain some insight into the reason why the imaginary part of the 
$f_{\rm s}$-mode fails to follow a universal scaling which depends
solely on the mass and radius of the star, we provide here a qulitative 
understanding based on the multipole formulas of the 
gravitational-wave luminosity for a single oscillation mode with 
frequency $\omega$ \citep{thorne1980}:
\begin{equation} 
{ {d E} \over {dt} } = \sum_{l=2}^{\infty} N_l \omega^{2 l + 2} \left(
| \delta D_{lm} |^2 + | \delta J_{lm} |^2 \right) ,
\end{equation}
where $N_l$ is some constant depending on the spherical harmonic index 
$l$. $\delta D_{lm}$ and $\delta J_{lm}$ are respectively the mass and 
current multipoles associated to the oscillation modes.
For a nonrotating two-fluid star with weak internal gravity 
(i.e., a Newtonian source), the multipoles are given by
\citep{andersson-2008}  
\begin{eqnarray}
\delta D_{lm} &=& \int ( \delta \rho_{\rm n} + \delta \rho_{\rm p} ) 
r^l Y_{lm}^* d V , \cr
&& \cr
\delta J_{lm} &=& {2 \over c} \sqrt{ l \over {l + 1} } 
\int r^l \left( \rho_{\rm n} \delta {\bf v}_{\rm n} + \rho_{\rm p} 
\delta {\bf v}_{\rm p} \right) \cdot {\bf Y}_{lm}^{B *} d V , 
\end{eqnarray}
where ${\bf Y}_{lm}^{B} = \left[ l(l+1)\right]^{-1/2} 
{\bf \hat r} \times \nabla Y_{lm}$ is the magnetic-type vector spherical 
harmonics.

The ordinary $f_{\rm o}$-mode of a two-fluid star is 
characterized by the fact that the two fluids are comoving, which implies
a large total density variation $\delta \rho = \delta \rho_{\rm n}
+ \delta \rho_{\rm p}$. Hence, similar to the $f$-mode of a single-fluid star,
it is expected that the gravitational-wave emission of a $f_{\rm o}$-mode 
is dominated by the mass multipole $\delta D_{lm}$. The damping time of the 
$l=2$ $f$-mode can be estimated by 
$\tau \sim E_m / \omega^{6} | \delta D_{22} |^2$. The mode energy 
$E_m$ can be calculated by giving the eigenfunction of the mode.  
For a homogenous single-fluid model, it can be shown that the damping timescale
depends only on the mass and radius of the star \citep{detweiler_1975}. 
In fact, the proposal for the leading scaling term $C^4$ in 
Eq.~(\ref{eq:nils_img_f}) was based on this rough estimation. 
Since the $f_{\rm o}$-mode of a two-fluid star is essentially the same as 
the standard $f$-mode, it is thus not surprising that the $f_{\rm o}$-mode 
also follows the same universal scaling law as we have 
seen in the numerical data. 

\begin{figure}
\centering
\includegraphics*[width=6.8cm]{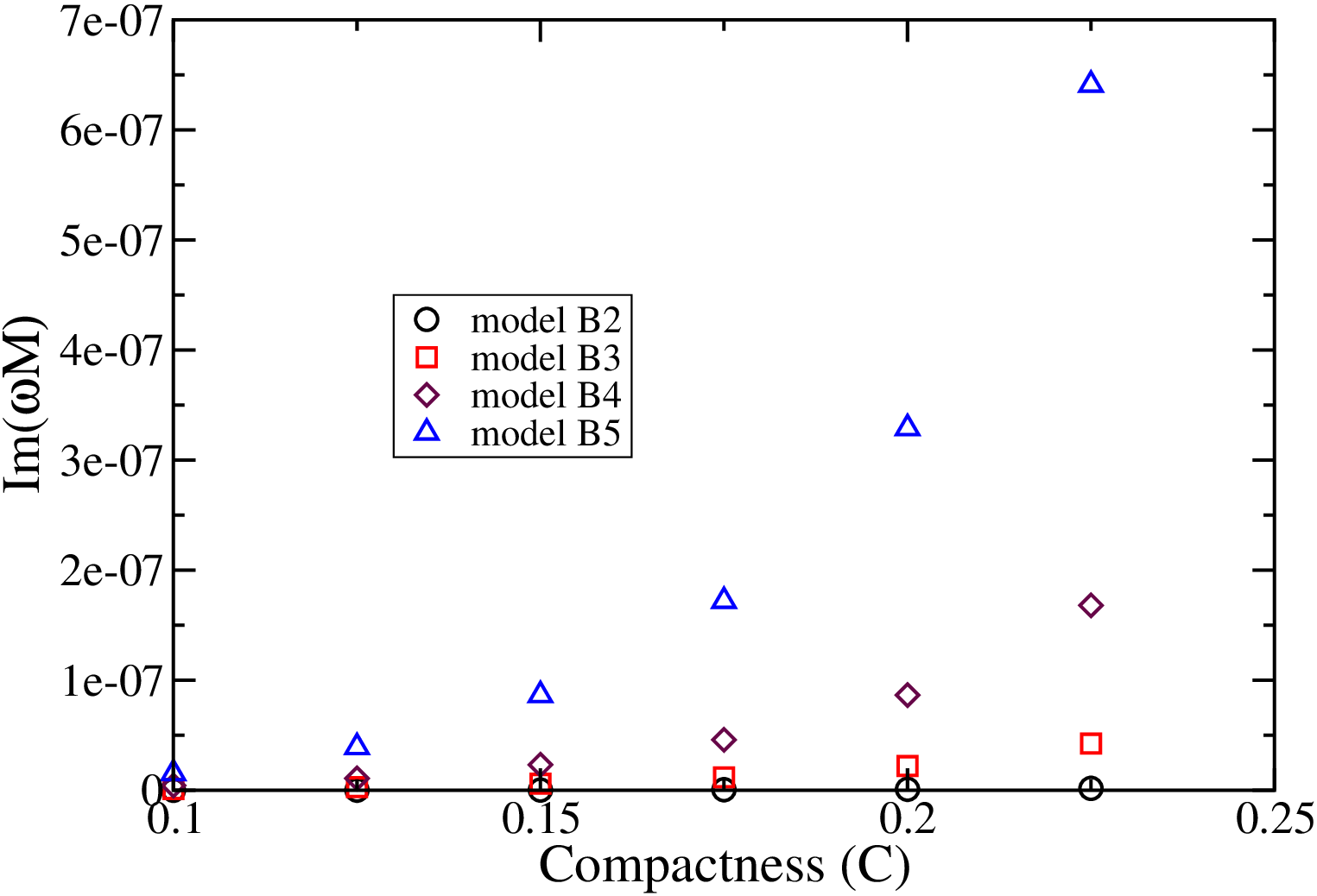}
\caption{${\rm Im}(\omega M)$ of the $f_{\rm s}$-modes
vs $C$ for the EOS 
models B$i$ (with $i=2 - 5$).}
\label{fig:B_fs_IM}
\end{figure}

\begin{figure}
\centering
\includegraphics*[width=6.8cm]{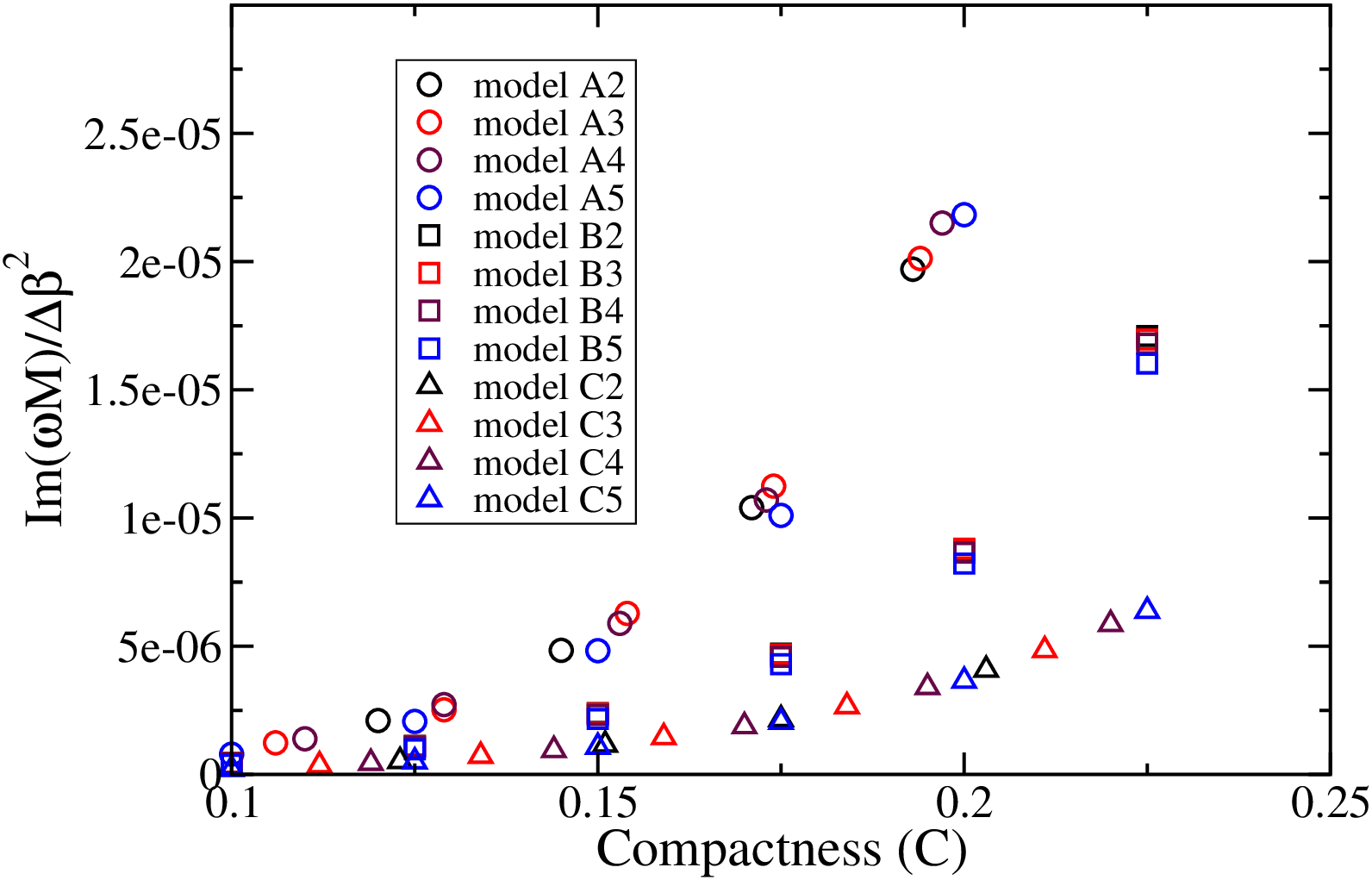}
\caption{${\rm Im}(\omega M) /\Delta \beta^2$ of the $f_{\rm s}$-modes
vs $C$ for the EOS models A$i$, B$i$, and C$i$ (with $i=2-5$).}
\label{fig:sc_fs_IM}
\end{figure}

\begin{figure}
\centering
\includegraphics*[width=6.8cm]{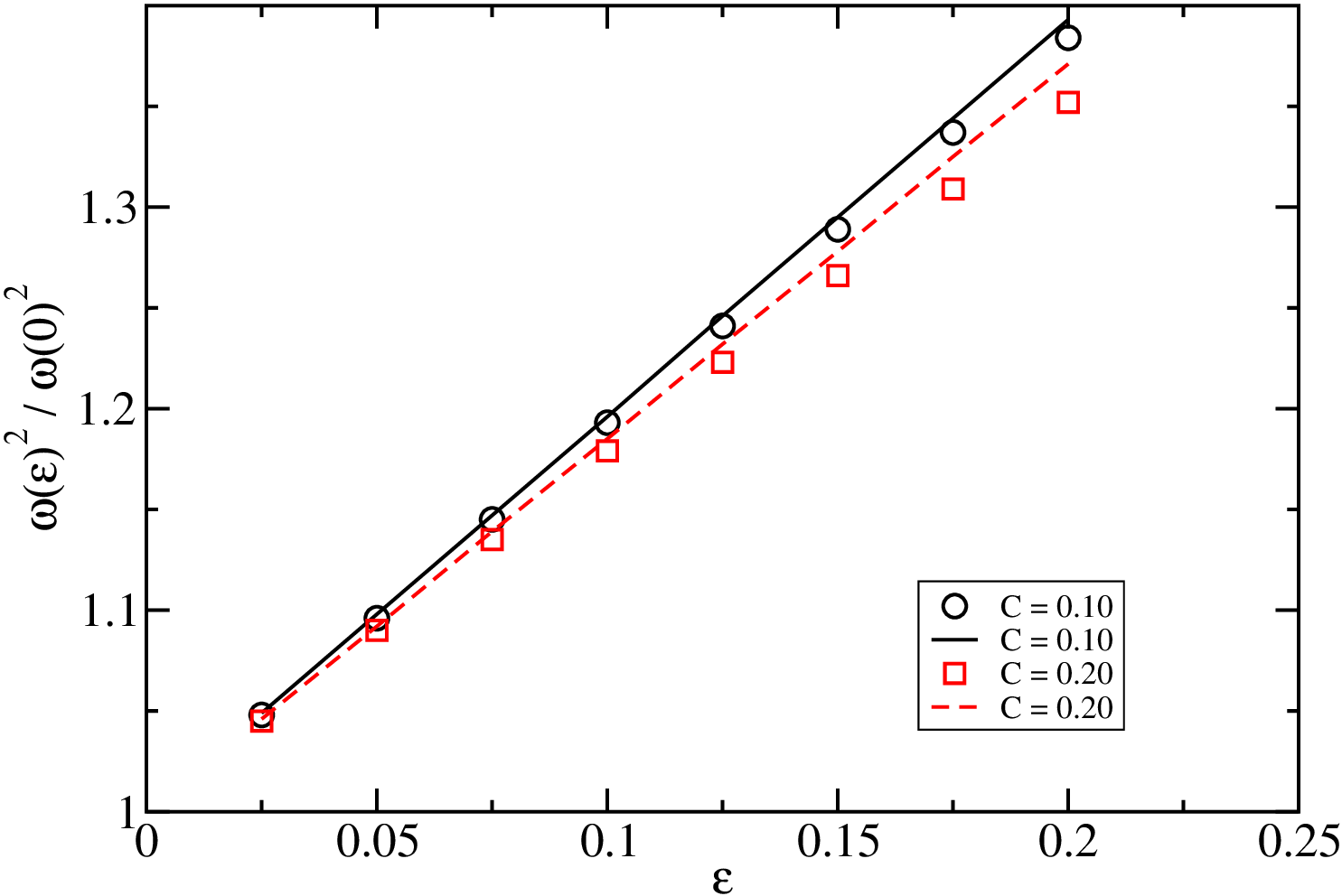}
\caption{$f_{\rm s}$-mode squared frequency $ \omega(\epsilon)^2 $ 
normalised by $\omega(0)^2$ (the squared frequency when there is no 
entrainment) versus $ \epsilon $ for model B2 at various compactness 
${\rm C}$. The circle and square symbols represent the numerical data.  
The solid and dashed lines represent the results of  
Eq.~(\ref{eq:w2_over_w2f2}). }
\label{fig:b2_w2_vs_e}
\end{figure}

\begin{figure}
\centering
\includegraphics*[width=6.8cm]{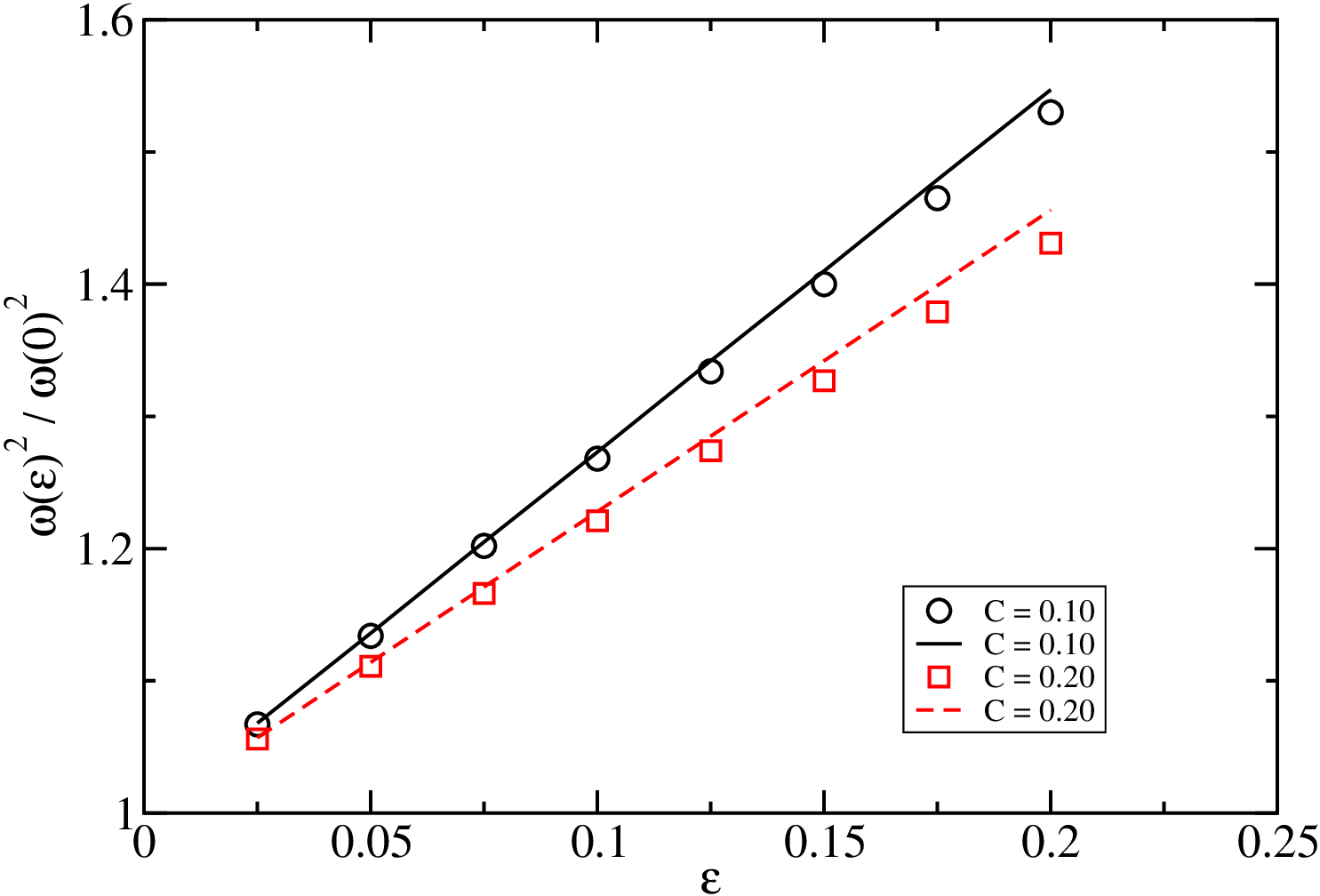}
\caption{Simliar to Figure~\ref{fig:b2_w2_vs_e}, but for model B5. }
\label{fig:b5_w2_vs_e}
\end{figure}

\begin{figure}
\centering
\includegraphics*[width=6.8cm]{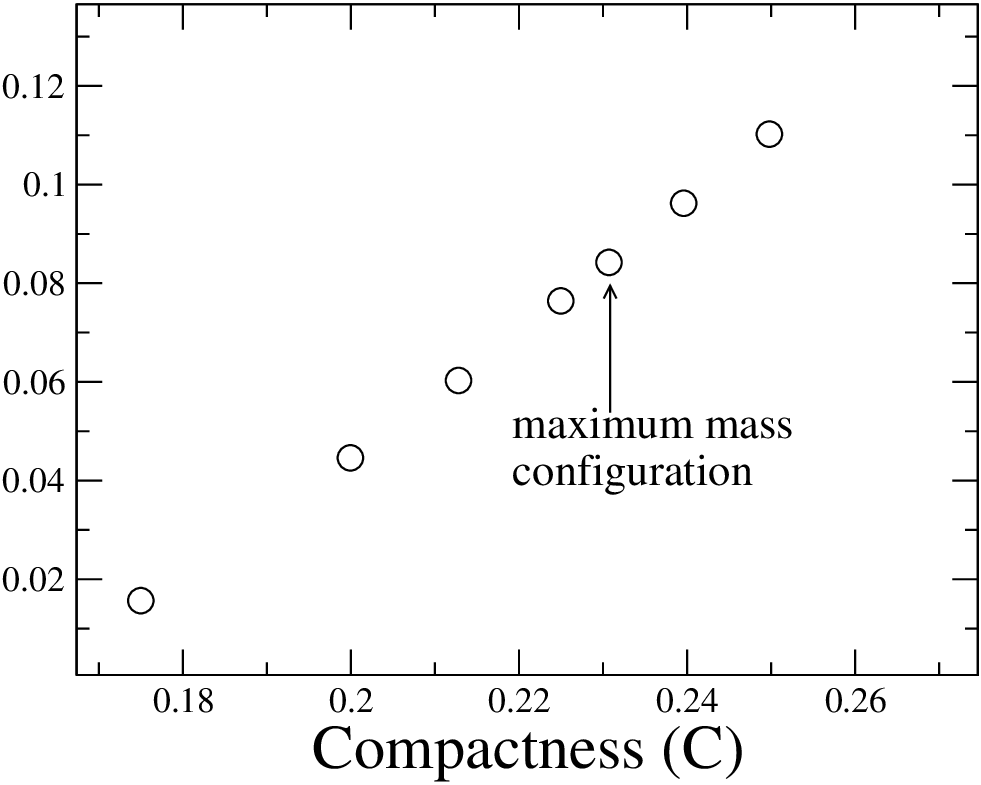}
\caption{Relative difference ($|\omega_{\rm num} - \omega_{\rm ana}|
/ \omega_{\rm ana}$) between the $f_{\rm s}$-mode frequency obtained by the 
relativistic numerical code $\omega_{\rm num}$
and the Newtonian analytic result~(\ref{eq:newton_fs_derive}) 
$\omega_{\rm ana}$ versus the compactness $C$ for model C5. }
\label{fig:modelC5_rel_diff}
\end{figure}

\begin{figure*}
\centering
\includegraphics*[width=13.6cm]{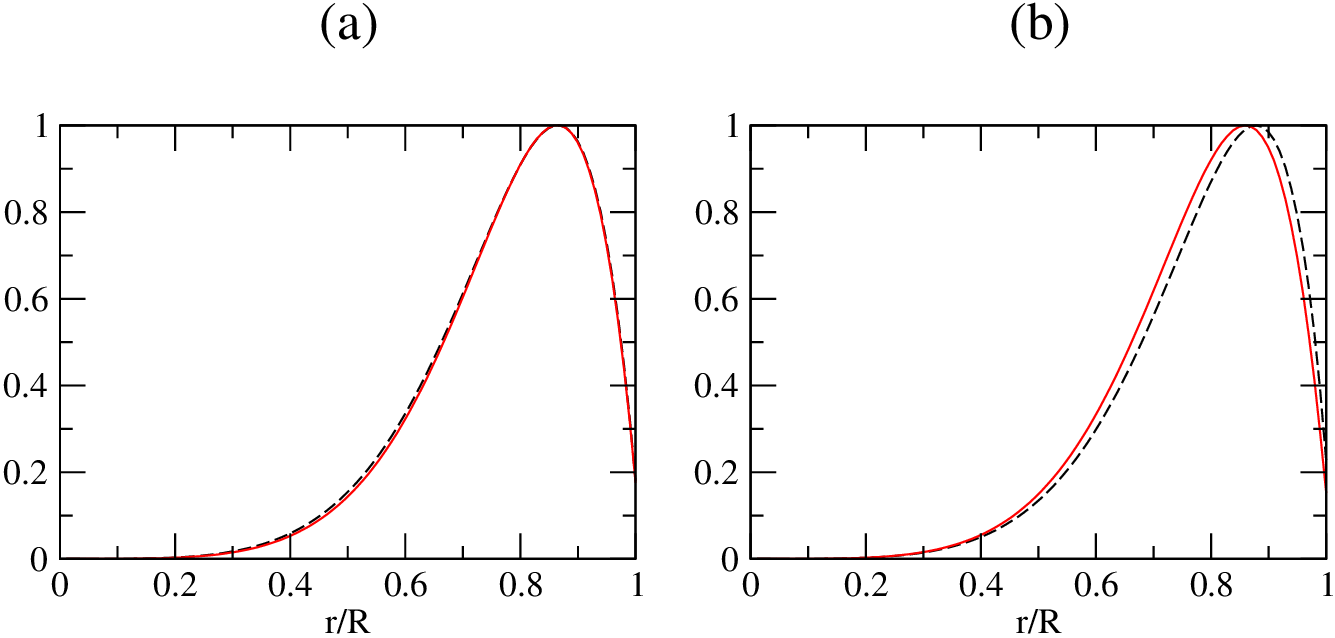}
\caption{Plots of $f_1(r)$ (dashed lines) and $f_2(r)$ (solid lines) 
as defined in Eq.(\ref{eq:f1f2}) against $r/R$ for (a) model B2 and (b) 
model B5 at $C = 0.15$. 
Both functions are normalised by their maximum values.}
\label{fig:B2B5term}
\end{figure*}

\begin{figure}
\centering
\includegraphics*[width=6.8cm]{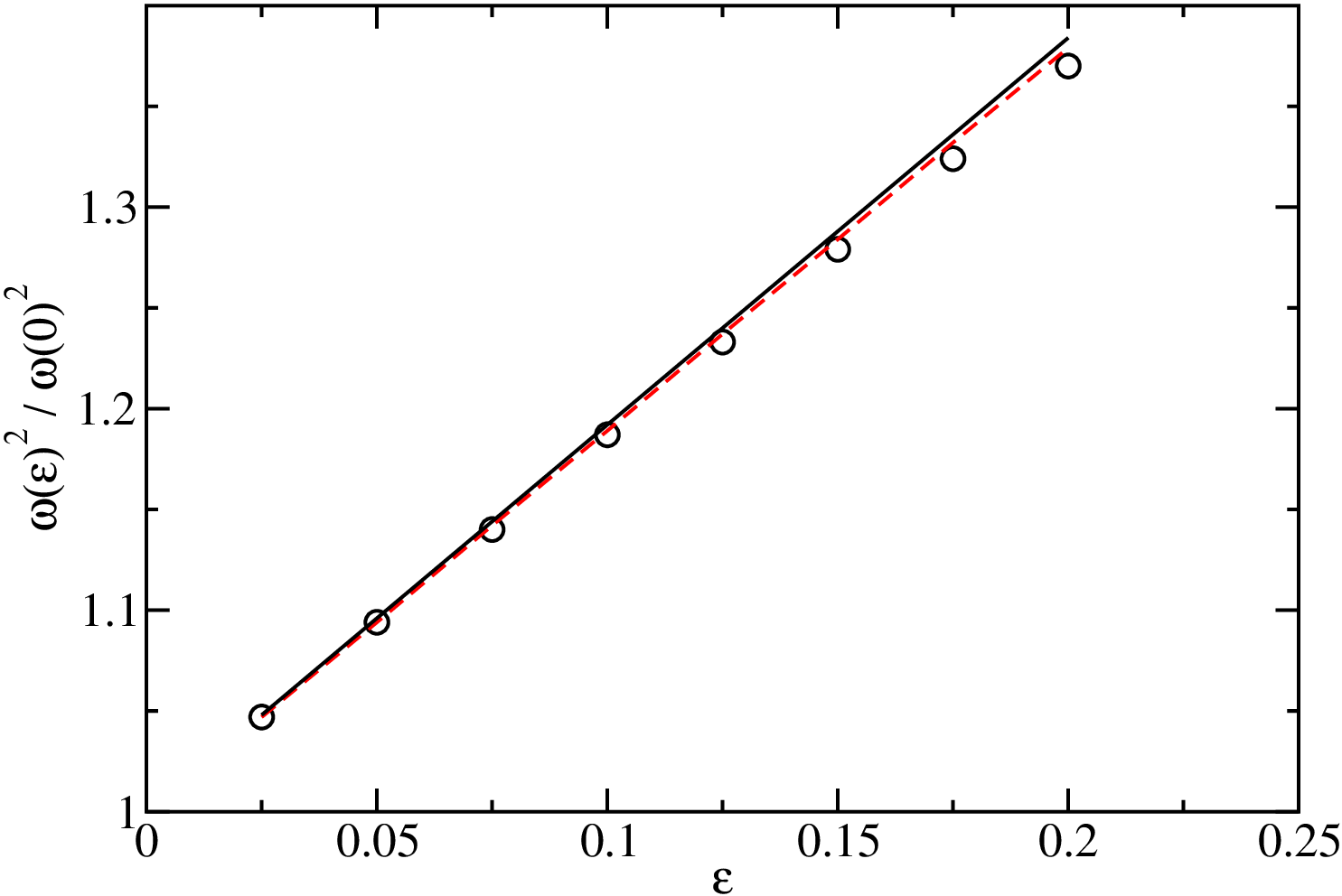}
\caption{$f_{\rm s}$-mode squared frequency $ \omega(\epsilon)^2 $ 
normalised by $\omega(0)^2$ (i.e., its value when there is no entrainment)
versus $ \epsilon $ for model B2 at compactness ${\rm C} = 0.15$. 
The circle data points are obtained directly from the relativistic numerical 
code. The solid and dashed lines are obtained respectively from 
Eqs.~(\ref{eq:1st_order_rel}) and (\ref{eq:1st_order_rel_final}). }
\label{fig:b2_w2_vs_e2}
\end{figure}

Now let us turn to the superfluid $f_{\rm s}$-mode. This mode is characterized
by the counter-moving motion of the two fluids in such a way that the total 
density variation, and hence the mass multipole, nearly vanish. 
It is thus conceivable that the current multipole $\delta J_{lm}$ could 
provide the main radiation mechanism for the $f_{\rm s}$-mode. It is also 
interesting to note that, for the case of two nearly symmetric fluids
(e.g., $\sigma_{\rm n}\approx\sigma_{\rm p}$ and 
$\beta_{\rm n} \approx\beta_{\rm p}$ in the master 
function~(\ref{eq:master_poly})), the superfluid modes become non-radiating 
as the current multipole nearly vanishes. 
This explains why the imaginary part of the mode complex frequency
${\rm Im}(\omega M) \rightarrow 0$ in the limit 
$\Delta \beta \rightarrow 0$ as shown in Figure~(\ref{fig:B_fs_IM}). 
In the general situation, however, the mass current
$\rho_{\rm n}\delta {\bf v}_{\rm n} + \rho_{\rm p}\delta {\bf v}_{\rm p}$
would depend on the mass fractions of the two components, which are determined 
by the condition of chemical equilibrium. 
The equilibrium condition in turn depends sensitively on the underlying 
EOS models and thus the damping timescale of the $f_{\rm s}$-mode would not 
have a simple scaling relation with the compactness of the star. 
Stellar models with the same global parameters (e.g., the compactness) could 
have vastly different fractions of the two particle species.  
The counter-moving character of the superfluid $f_{\rm s}$-mode would then 
lead to different amount of gravitational-wave emission among the stellar
models.

Finally, we note that it might be possible to perform a more detailed
quantitative analysis by expanding the current multipole and extract the 
leading dependence of the damping time of the $f_{\rm s}$-mode on the local 
thermodynamics quantities. Such kind of analysis might explain the relation
(\ref{eq:im_fsf1}) seen in the numerical data. 
A natural starting point would be the extension of the recent work of 
\citet{andersson-2008}, in which the dependence of the damping time of the 
ordinary $f_{\rm o}$-mode due to the so-called mutual friction on the 
thermodynamics quantities has been studied.


\bibliography{biblio}

\newcommand{\noopsort}[1]{} \newcommand{\printfirst}[2]{#1}
  \newcommand{\singleletter}[1]{#1} \newcommand{\switchargs}[2]{#2#1}
\begin{thebibliography}{46}
\expandafter\ifx\csname natexlab\endcsname\relax\def\natexlab#1{#1}\fi

\bibitem[{Abbott {et~al.}(2007{\natexlab{a}})}]{abbott-2007-1}
Abbott, B., {et~al.} 2007{\natexlab{a}}, \apj, 659, 918

\bibitem[{Abbott {et~al.}(2007{\natexlab{b}})}]{abbott-2007-2}
---. 2007{\natexlab{b}}, \prd, 76, 042001

\bibitem[{Abbott {et~al.}(2008)}]{abbott-2008sgr}
---. 2008, Phys. Rev. Lett., 101, 211102

\bibitem[{Alford \& Reddy(2003)}]{alford-2003-67}
Alford, M., \& Reddy, S. 2003, \prd, 67, 074024

\bibitem[{Alford(2004)}]{alford-2004-30}
Alford, M.~G. 2004, J. Phys. G, 30, S441

\bibitem[{Andersson(2003)}]{andersson2003gwi}
Andersson, N. 2003, Class. Quantum Grav., 20, R105

\bibitem[{Andersson \& Comer(2001{\natexlab{a}})}]{andersson2001dsn}
Andersson, N., \& Comer, G.~L. 2001{\natexlab{a}}, Mon. Not. R. Astron. Soc.,
  328, 1129

\bibitem[{Andersson \& Comer(2001{\natexlab{b}})}]{andersson2001srg}
---. 2001{\natexlab{b}}, Class. Quantum Grav., 18, 969

\bibitem[{Andersson \& Comer(2007)}]{Andersson_Comer_review}
---. 2007, Living Rev. Rel., 10, 1

\bibitem[{Andersson {et~al.}(2005)Andersson, Comer, \&
  Glampedakis}]{andersson-2005-763}
Andersson, N., Comer, G.~L., \& Glampedakis, K. 2005, Nucl. Phys. A, 763, 212

\bibitem[{Andersson {et~al.}(2002)Andersson, Comer, \&
  Langlois}]{andersson-2002-66}
Andersson, N., Comer, G.~L., \& Langlois, D. 2002, \prd, 66, 104002

\bibitem[{Andersson {et~al.}(2008)Andersson, Glampedakis, \&
  Haskell}]{andersson-2008}
Andersson, N., Glampedakis, K., \& Haskell, B. 2008, arXiv:0812.3023 [astro-ph]

\bibitem[{Andersson \& Kokkotas(1996)}]{andersson96}
Andersson, N., \& Kokkotas, K.~D. 1996, Phys. Rev. Lett., 77, 4134

\bibitem[{Andersson \& Kokkotas(1998)}]{andersson1998tgw}
---. 1998, Mon. Not. R. Astron. Soc., 299, 1059

\bibitem[{Benhar {et~al.}(1999)Benhar, Berti, \& Ferrari}]{Benhar:1998au}
Benhar, O., Berti, E., \& Ferrari, V. 1999, Mon. Not. R. Astron. Soc., 310, 797

\bibitem[{Benhar {et~al.}(2004)Benhar, Ferrari, \& Gualtieri}]{benhar04}
Benhar, O., Ferrari, V., \& Gualtieri, L. 2004, \prd, 70, 124015

\bibitem[{Campolattaro \& Thorne(1970)}]{campolattaro1970npg}
Campolattaro, A., \& Thorne, K.~S. 1970, \apj, 159, 847

\bibitem[{Carter(1989)}]{carter1989rfd}
Carter, B. 1989, in Lecture Notes in Mathematics, Vol. 1385, Relativistic Fluid
  Dynamics (Noto, 1987), ed. A.~Anile \& M.~Choquet-Bruhat (Heidelberg,
  Germany: Springer-Verlag), 1--64

\bibitem[{Carter \& Langlois(1995)}]{carter-1995-454}
Carter, B., \& Langlois, D. 1995, Nucl. Phys. B, 454, 402

\bibitem[{Chandrasekhar(1981)}]{chandra81}
Chandrasekhar, S. 1981, Hydrodynamics and hydromagnetic stability (New York:
  Dover)

\bibitem[{Comer \& Langlois(1993)}]{comer1993hfm}
Comer, G.~L., \& Langlois, D. 1993, Class. Quantum Grav., 10, 2317

\bibitem[{Comer \& Langlois(1994)}]{comer1994hfr}
---. 1994, Class. Quantum Grav., 11, 709

\bibitem[{Comer {et~al.}(1999)Comer, Langlois, \& Lin}]{comer-1999-60}
Comer, G.~L., Langlois, D., \& Lin, L.~M. 1999, \prd, 60, 104025

\bibitem[{Detweiler(1975)}]{detweiler_1975}
Detweiler, S.~L. 1975, \apj, 197, 203

\bibitem[{Detweiler \& Ipser(1973)}]{detweiler_1973}
Detweiler, S.~L., \& Ipser, J.~R. 1973, \apj, 185, 685

\bibitem[{Ferrari \& Gualtieri(2008)}]{ferrari08}
Ferrari, V., \& Gualtieri, L. 2008, Gen. Rel. Grav., 40, 945

\bibitem[{Kokkotas {et~al.}(2001)Kokkotas, Apostolatos, \&
  Andersson}]{kokkotas01}
Kokkotas, K.~D., Apostolatos, T.~A., \& Andersson, N. 2001, Mon. Not. R.
  Astron. Soc., 320, 307

\bibitem[{Kokkotas \& Schmidt(1999)}]{kokkotas1999qnm}
Kokkotas, K.~D., \& Schmidt, B.~G. 1999, Living Rev. Rel., 2, 2

\bibitem[{Langlois {et~al.}(1998)Langlois, Sedrakian, \&
  Carter}]{langlois1998drr}
Langlois, D., Sedrakian, D.~M., \& Carter, B. 1998, Mon. Not. R. Astron. Soc.,
  297, 1189

\bibitem[{Lee(1995)}]{lee1995non}
Lee, U. 1995, \aap, 303, 515

\bibitem[{Lin {et~al.}(2008)Lin, Andersson, \& Comer}]{lin-2007}
Lin, L.-M., Andersson, N., \& Comer, G.~L. 2008, \prd, 78, 083008

\bibitem[{Lindblom \& Mendell(1994)}]{lindblom1994osn}
Lindblom, L., \& Mendell, G. 1994, \apj, 421, 689

\bibitem[{{Lombardo}(1999)}]{lombardo-1999}
{Lombardo}, U. 1999, in Nuclear Methods and the Nuclear Equation of State, ed.
  M.~{Baldo} (Singapore: World Scientific), 458

\bibitem[{Lombardo \& Schulze(2001)}]{lombardo-2001-578}
Lombardo, U., \& Schulze, H.~J. 2001, in Lecture Notes in Physics, Vol. 578,
  Physics of Neutron Star Interiors, ed. D.~Blaschke, N.~K. Glendenning, \&
  A.~Sedrakian (Berlin/Heidelberg: Springer), 30

\bibitem[{Lyne(1993)}]{lyne-1993}
Lyne, A.~G. 1993, Pulsars as Physics Laboratories, ed. R.~D. Blandford,
  A.~Hewish, A.~G. Lyne, \& L.~Mestel (Oxford: Oxford University Press), 29--38

\bibitem[{Price \& Thorne(1969)}]{price1969tnr}
Price, R., \& Thorne, K.~S. 1969, \apj, 155, 163

\bibitem[{Prix \& Rieutord(2002)}]{prix2002aon}
Prix, R., \& Rieutord, M. 2002, A\&A, 393, 949

\bibitem[{Radhakrishnan \& Manchester(1969)}]{Radhakrishnan-1969}
Radhakrishnan, V., \& Manchester, R.~N. 1969, Nature (London), 222, 228

\bibitem[{Thorne(1969{\natexlab{a}})}]{thorne1969npg1}
Thorne, K.~S. 1969{\natexlab{a}}, \apj, 158, 1

\bibitem[{Thorne(1969{\natexlab{b}})}]{thorne1969npg2}
---. 1969{\natexlab{b}}, \apj, 158, 997

\bibitem[{Thorne(1980)}]{thorne1980}
---. 1980, Rev. Mod. Phys., 52, 299

\bibitem[{Thorne \& Campolattaro(1967)}]{thorne1967tnr}
Thorne, K.~S., \& Campolattaro, A. 1967, \apj, 149, 591

\bibitem[{Tsui \& Leung(2005{\natexlab{a}})}]{tsui2005prl}
Tsui, L.~K., \& Leung, P.~T. 2005{\natexlab{a}}, Phys. Rev. Lett., 95, 151101

\bibitem[{Tsui \& Leung(2005{\natexlab{b}})}]{tsui2005uqn}
---. 2005{\natexlab{b}}, Mon. Not. R. Astron. Soc., 357, 1029

\bibitem[{Wong(2008)}]{wong_thesis08}
Wong, K.-S. 2008, Master thesis, The Chinese University of Hong Kong

\bibitem[{Yoshida \& Lee(2003)}]{yoshida03}
Yoshida, S., \& Lee, U. 2003, \prd, 67, 124019

\end{thebibliography}

\clearpage

\begin{deluxetable}{c c c c c}
 \tablecaption{Models for the ``polytropic'' EOS defined in 
      equation~(\ref{eq:master_poly})}
 \tablehead{ 
   \colhead{Model\tablenotemark{a}}
     & \colhead{$\beta_n$} 
     & \colhead{$\beta_p$} 
     & \colhead{$\bar{\beta}$\tablenotemark{b}} 
     & \colhead{$\Delta\beta$\tablenotemark{c}} 
}    
 \startdata
   A1 & 1.9 & 1.9 & 1.9 & 0.0  \\
   A2 & 1.905 & 1.895 & 1.9 & 0.01 \\
   A3 & 1.925 & 1.875 & 1.9 & 0.05 \\
   A4 & 1.95 & 1.85 & 1.9 & 0.1  \\
   A5 & 2.0 & 1.8 & 1.9 & 0.2  \\
   \hline
   B1 & 2.0 & 2.0 & 2.0 & 0.0 \\
   B2 & 2.005 & 1.995 & 2.0 & 0.01 \\
   B3 & 2.025 & 1.975 & 2.0 & 0.05 \\
   B4 & 2.05 & 1.95 & 2.0 & 0.1 \\
   B5 & 2.1 & 1.9 & 2.0 & 0.2 \\
   \hline
   C1 & 2.1 & 2.1 & 2.1 & 0.0  \\
   C2 & 2.105 & 2.095 & 2.1 & 0.01 \\
   C3 & 2.125 & 2.075 & 2.1 & 0.05 \\
   C4 & 2.15 & 2.05 & 2.1 & 0.1 \\
   C5 & 2.2 & 2.0 & 2.1 & 0.2  
   \enddata
\tablenotetext{a}{The parameters $\sigma_n=\sigma_p=0.5 m$ are fixed in all 
models.}
\tablenotetext{b}{$\bar{\beta}=(\beta_n+\beta_p)/2$.}
\tablenotetext{c}{$\Delta\beta=\beta_n-\beta_p$. \\}
\label{tab:1}
\end{deluxetable}

\begin{table}[tbp]
\caption{Values of the ratio $\eta$ as defined in Eq.(\ref{eq:ratio_ud}) 
for the $f_{\rm s}$-modes of models B2 and B5 (see Table~\ref{tab:1})
at various compactness $C$}
\label{tab:U}
\begin{center}
\begin{tabular}{|c||c|c|}
\hline
Model & ${\rm C}$ & $\eta$ \\
\hline
\multirow{3}{*}{B2} & $0.10$ & $1.89 \times 10^{-5}$\\
						   & $0.15$ & $2.24 \times 10^{-5}$\\
                           & $0.20$ & $2.83 \times 10^{-5}$\\
\hline 
\multirow{3}{*}{B5} & $0.10$ & $6.95 \times 10^{-3}$\\
						   & $0.15$ & $8.39 \times 10^{-3}$\\
                           & $0.20$ & $1.11 \times 10^{-2}$\\
\hline
\end{tabular}
\end{center}
\end{table}

\clearpage

\end{document}